\documentclass[journal]{IEEEtran}

\usepackage{booktabs}
\usepackage{amsmath}
\usepackage{amsfonts}
\usepackage{xcolor}
\usepackage{pifont}
\usepackage{multirow}
\usepackage{graphicx}
\usepackage{subcaption}
\usepackage{microtype}
\usepackage[colorlinks=true,citecolor=blue]{hyperref}

\definecolor{deepblue}{rgb}{0, 0, 0.9}
\definecolor{lightgray}{rgb}{0.4, 0.4, 0.4}

\newcommand{\cmark}{\ding{51}} 
\newcommand{\xmark}{\ding{55}} 

\begin{document}

\title{Why Pre-trained Models Fail: Feature Entanglement in Multimodal Depression Detection}
\author{Xiangyu Zhang, Beena Ahmed~\IEEEmembership{Member,~IEEE,} Julien Epps~\IEEEmembership{Senior Member,~IEEE} 
\thanks{Xiangyu Zhang, Beena Ahmed, Julien Epps are with the School of Electrical Engineering and Telecommunications, University of New South Wales, Sydney, Australia.}
}  

\maketitle
\begin{abstract}
Depression remains a pressing global mental health issue, driving considerable research into AI-driven detection approaches. While pre-trained models, particularly speech self-supervised models (SSL Models), have been applied to depression detection, they show unexpectedly poor performance without extensive data augmentation. Large Language Models (LLMs), despite their success across various domains, have not been explored in multi-modal depression detection. In this paper, we first establish an LLM-based system to investigate its potential in this task, uncovering fundamental limitations in handling multi-modal information. Through systematic analysis, we discover that the poor performance of pre-trained models stems from the conflation of high-level information, where \textcolor{black}{one of the important reasons is that} high-level features derived from both content and speech are mixed within pre-trained models model representations, making it challenging to establish effective decision boundaries. To address this, we propose an information separation framework that disentangles these features, significantly improving the performance of both SSL models and LLMs in depression detection. Our experiments validate this finding and demonstrate that the integration of separated features yields substantial improvements over existing approaches, providing new insights for developing more effective multi-modal depression detection systems.
\end{abstract}

\section{Introduction}
Depression is a widespread mental health disorder, impacting 10-15\% of the global population and marked by persistent low mood, diminished interest and fatigue, making it a significant and costly health challenge~\cite{walker2018prevalence}. Traditional methods for diagnosing and treating depression are often resource-intensive and may lack efficacy, leading researchers to increasingly focus on developing automated systems for depression detection. Pre-trained models, particularly speech self-supervised models (SSL Models), have emerged as promising tools for this task, given their ability to learn from limited labeled data~\cite{baevski2020wav2vec,chen2022wavlm}. However, these models consistently show poor performance in multi-modal depression detection without extensive data augmentation~\cite{wu2023self,bao2023somatisation}, raising fundamental questions about their limitations in this critical healthcare application.

This unexpected failure of pre-trained models in depression detection leads to our primary research questions: (1) Why do SSL models, despite their proven effectiveness across various speech tasks, perform poorly in depression detection? (2) What are the fundamental challenges in combining speech and text modalities for depression detection? (3) How can we design a system that effectively addresses these limitations without relying on extensive data augmentation?
\begin{figure}[t]
    \centering
    \includegraphics[width=0.5\textwidth]{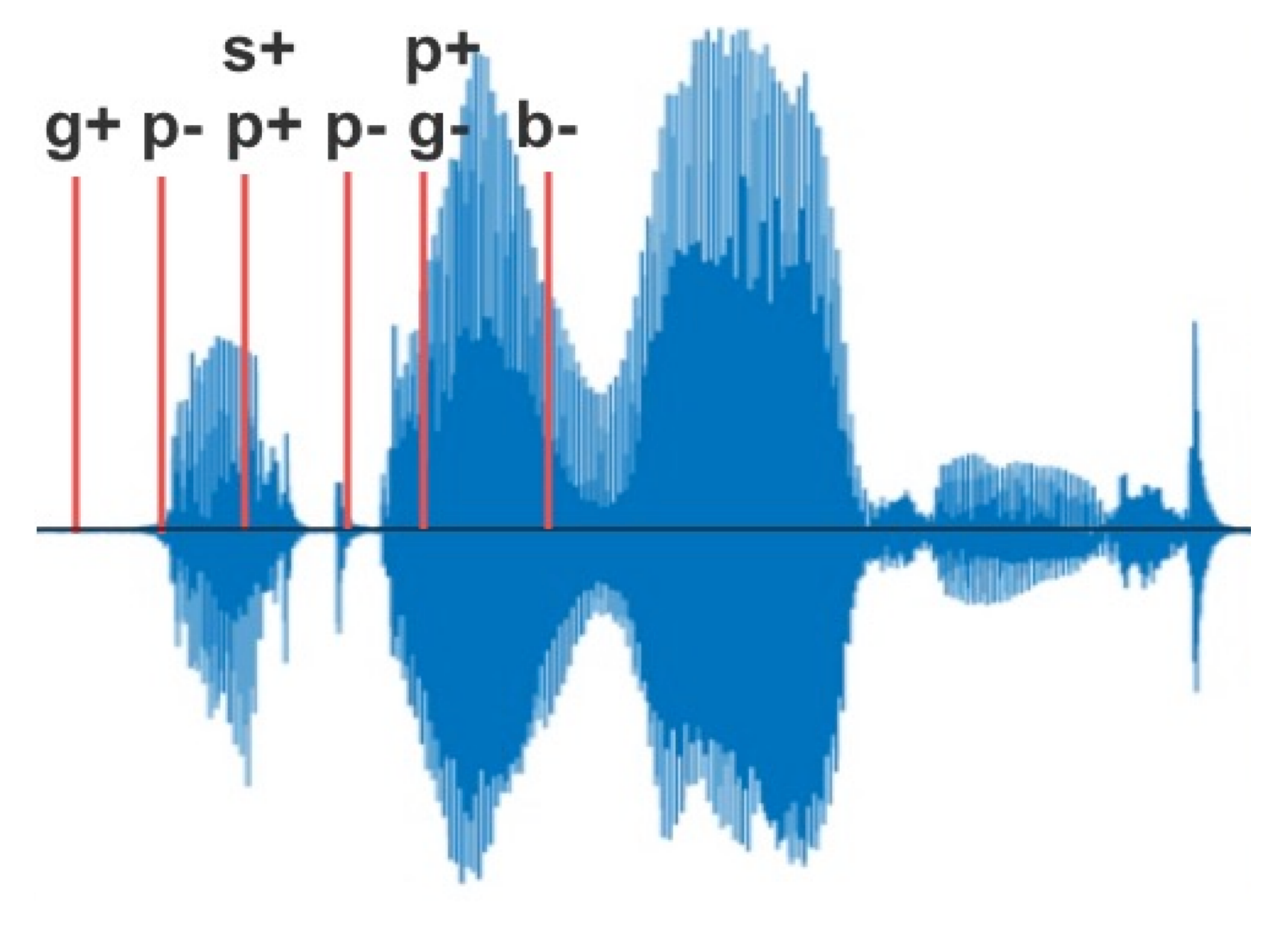}
    \caption{Example of acoustic landmarks (2-gram landmarks (g+p-), (s+p+), (p-,p+), ..., (g-b-)). Landmarks can discretize speech into a series of tokens.}\vspace{-5mm}
    \label{fig:landmark}
\end{figure}

To systematically investigate these challenges, we turn to another type of pre-trained model - Large Language Models (LLMs), which have recently demonstrated unprecedented capabilities in understanding complex patterns and generalizing across domains~\cite{chowdhery2023palm,touvron2023llama}. As the most advanced pre-trained models to date, LLMs have shown remarkable success in various healthcare applications~\cite{oh2023chatgpt,wang2023can,lahat2023evaluating} and could potentially capture subtle linguistic patterns that might indicate depression. To enable this investigation, we develop a system that integrates speech information through acoustic landmarks - discrete markers capturing temporal speech patterns~\cite{liu1996landmark, zhang2024auto, stevens2002toward}. These landmarks provide an efficient alternative to memory-intensive continuous representations, making them well-suited for integration with LLMs.
\begin{figure*}[t]
    \centering
    \includegraphics[width=1\textwidth]{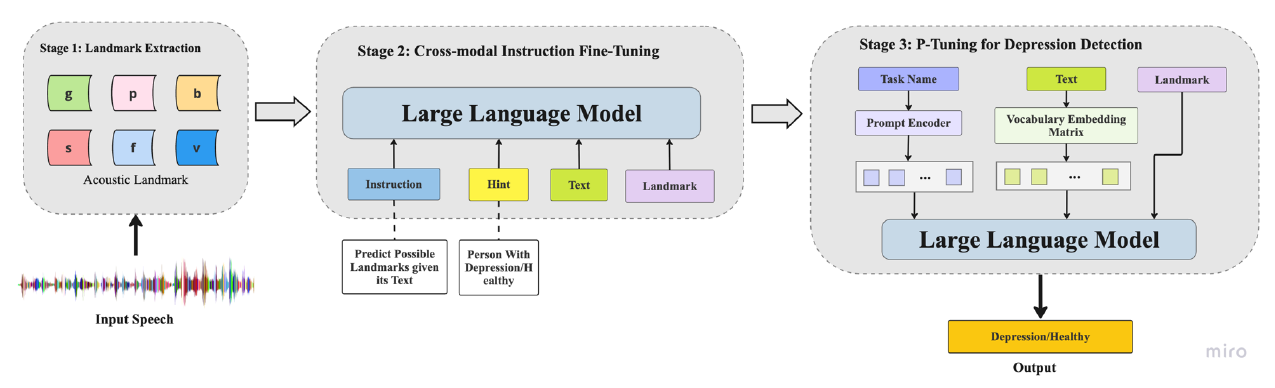}
    \caption{Overview of the proposed LLM-landmark depression detection baseline system, broadly categorized into three stages: landmark detection (on the left), cross-modal instruction fine-tuning (in the middle), and P-tuning for depression detection (on the right).} \vspace{-5mm}
    \label{fig:overview}
\end{figure*}


However, our systematic investigation reveals that even these powerful models struggle with multi-modal depression detection, suggesting a more fundamental issue than model capacity or architecture. This observation leads to our core hypothesis: the challenge lies in how these pre-trained models internally represent and process multi-modal information, \textcolor{black}{where the conflation of high-level features derived from both content and speech creates confused decision boundaries, representing one of the important contributing factors to SSL model limitations in depression detection}. Traditional approaches have attempted to improve performance through extensive data augmentation~\cite{wu2023self,zhang2024llms}, but this workaround fails to address the underlying problem and contradicts the intended purpose of pre-trained models - to generalize effectively with limited labeled data.

Based on this hypothesis, we propose an information separation framework that deliberately disentangles content-specific and speech-specific features. This approach represents a fundamental shift from previous methods that attempt to combine modalities directly. Our experimental results validate this hypothesis, demonstrating that separated, specialized representations significantly outperform traditional approaches without relying on data augmentation. Moreover, by integrating our speech dense vectors (designed to capture only speech-specific information) with the LLM embeddings, we achieve substantial improvements in LLM-based depression detection system. This result confirms that while content and speech information in speech SSL Models should be processed separately, they can be highly complementary when properly integrated, leading to good performance in multi-modal depression detection.

The remainder of this paper is organized as follows. Section~\ref{baseline system} presents our initial LLM-based system we published previously~\cite{zhang2024llms} and its limitations, Section~\ref{Limitation} analyzes why pre-trained models fail in depression detection, Section~\ref{new system} introduces our information separation framework, and Section~\ref{result},\ref{LLM SECTION} details the experimental results and analysis.

\section{Related Work}
\subsection{Acoustic Landmarks}
The concept of acoustic landmarks has its roots in studies of distinctive features~\cite{garvin1953preliminaries,zhang2024auto}. Some researchers suggest that for certain phonetic distinctions, listeners depend on acoustic landmarks to capture essential cues necessary for interpreting the associated distinctive features~\cite{liu1996landmark}. This view underscores the significance of landmarks in auditory perception and speech comprehension. Over time, acoustic landmarks have found applications not only in speech recognition~\cite{liu1996landmark,he2019ctc} but also in addressing issues related to mental health~\cite{huang2018depression, huang2019investigation}. Although definitions of acoustic landmarks vary somewhat across studies, Boyce et al.\cite{boyce2012speechmark} extended Liu’s earlier work\cite{liu1996landmark} by introducing a MATLAB-based landmark detection toolkit (SpeechMark), which has since become one of the most widely utilized resources for landmark technology.

\subsection{Depression Detection}
AI technology has been employed in automatic depression detection for several years, beginning with traditional approaches like Support Vector Machines (SVMs)~\cite{noble2006support}, as investigated by Cummins and others~\cite{cummins2011investigation, huang2018depression, huang2019investigation}. With advancements in deep learning~\cite{gulati2020conformer,zhang2024mamba}, researchers have increasingly adopted deep learning methods for this task. Zhao et al. have explored transformer models for processing speech in depression detection~\cite{zhao2020hierarchical}, while Shen and colleagues utilized BI-LSTM architectures to integrate text and speech inputs~\cite{shen2022automatic}. Building on these methods, Wu~\cite{wu2023self} implemented speech self-supervised models~\cite{chen2022wavlm, hsu2021hubert, efficient_lid} and integrated them with RoBERTa~\cite{liu2019roberta}, creating a more comprehensive multimodal framework that combines both text and audio for depression detection.

\subsection{Parameter-Efficient Fine-tuning}
Parameter-efficient fine-tuning (PEFT) methods have gained significant attention as alternatives to full model fine-tuning, particularly for large pre-trained models. LoRA~\cite{hu2022lora} introduces low-rank adaptation matrices to modify key weight matrices during fine-tuning while freezing the original model parameters. Prefix tuning~\cite{li2021prefix} prepends trainable continuous tokens to the input, allowing task-specific adaptation without modifying the model architecture. P-tuning~\cite{LIU2023} extends this concept by introducing trainable prompts at different layers of the model. These approaches have demonstrated comparable performance to full fine-tuning while requiring significantly fewer trainable parameters. For multimodal tasks, BitFit~\cite{zaken2022bitfit} showed that adapting only bias terms can effectively transfer pre-trained models across modalities.

\textcolor{black}{\subsection{Emotion and Affective Computing in Depression Detection}}

\textcolor{black}{Depression, as an affective disorder, is fundamentally characterized by persistent alterations in emotional states, including prolonged sadness, diminished interest, and emotional numbing~\cite{montejo2024survey}. The two core features of depression are depressed mood and anhedonia, where anhedonia represents a reduced ability to experience pleasure or diminished interest in previously enjoyable activities~\cite{carlsson2019distress}. Understanding how these emotional manifestations are expressed and can be detected across different modalities is crucial for developing effective automated depression detection systems~\cite{nasir2019multimodal,shen2022automatic}. The multimodal nature of emotional expression makes depression detection particularly challenging, as emotional cues manifest differently in speech and text modalities, requiring sophisticated approaches to capture and integrate these complementary sources of affective information~\cite{fan2019multi,mamidisetti2022multimodal}.}


\textcolor{black}{\textbf{Emotional Expression in Speech.} Speech naturally carries rich emotional information through multiple acoustic dimensions. Prosodic features such as fundamental frequency variations, speaking rate, and pause patterns can reveal emotional states~\cite{hashem2023speech}. For example, individuals with depression often exhibit flattened prosody, reduced vocal energy, longer pauses, and slower speaking rates~\cite{fairbairn2013detecting,seifpanahi2023association}. Spectral characteristics, including formant frequencies and spectral energy distribution, further encode emotional content~\cite{low2020acoustic}. These acoustic manifestations provide direct physiological indicators of emotional states that are difficult to consciously control or mask~\cite{yang2017bio}.}

\textcolor{black}{\textbf{Limitations of Text-Based Emotional Analysis.} While text processing has advanced significantly in emotion recognition~\cite{etienne-etal-2024-emotion,kim2019analysis}, the text modality faces inherent limitations in conveying emotional context. Text lacks the prosodic and paralinguistic cues that naturally accompany spoken language, making emotional interpretation heavily dependent on lexical content and linguistic structure~\cite{walker2018prevalence}. For instance, the phrase "I'm fine" could indicate genuine well-being or mask underlying distress, but without acoustic cues such as vocal tremor, hesitation, or monotone delivery, the true emotional state remains ambiguous~\cite{williamson2016detecting,wu2022novel}. Furthermore, individuals with depression may unconsciously modify their written expression to appear more positive, while their speech patterns reveal underlying emotional struggles through involuntary acoustic markers.}

\textcolor{black}{\textbf{Multimodal Emotional Cues in Depression.} Recent studies demonstrate that different textual contexts convey emotional states through distinct linguistic mechanisms, including word choice, sentence structure, and discourse patterns~\cite{kim2019analysis}. However, these textual indicators often require sophisticated contextual understanding and may be culturally or individually variable. In contrast, acoustic emotional markers tend to be more universal and physiologically grounded~\cite{liu1996landmark,zhang-etal-2025-speecht}. For example, depressed speech may exhibit reduced pitch variability and increased pause duration regardless of the spoken language or cultural background, while textual expressions of the same emotional state may vary significantly across linguistic and cultural contexts.}

\textcolor{black}{\textbf{Challenges in Multimodal Emotion Integration.} The intersection of emotion recognition and mental health detection has revealed significant challenges in effectively combining multimodal emotional cues~\cite{montejo2024survey}. Each modality captures complementary aspects of emotional expression: speech provides physiological and involuntary emotional markers, while text offers cognitive and deliberate emotional content. However, existing approaches to multimodal fusion often struggle to preserve the distinctive emotional characteristics of each modality, leading to suboptimal performance in depression detection tasks~\cite{wu2023self,zheng2023two}. This challenge motivates the need for more sophisticated approaches to multimodal integration that can effectively leverage the complementary emotional information from both speech and text while preserving their unique characteristics.}

\textcolor{black}{\subsection{Contextualizing Our Analytical Study within Recent Fusion Paradigms}}

\textcolor{black}{Recently, there has been a surge of advanced deep learning architectures designed for depression detection, predominantly focusing on sophisticated multimodal feature fusion strategies~\cite{10872825,li2025depressinstruct, cheng2025multisource, niu2024depressionmlp}. For example, cutting-edge studies have explored decoupled multi-perspective fusion networks~\cite{10872825} and large speech-language instruction tuning~\cite{li2025depressinstruct} to better align audio and textual modalities. Beyond speech and text, visual and temporal dynamics—such as facial keypoints and action units modeled via Multi-Layer Perceptrons (MLPs)—have also been effectively integrated~\cite{niu2024depressionmlp}, alongside fine-grained spatial-temporal attention mechanisms to capture cross-modal emotional cues~\cite{cheng2025multisource}.
While these state-of-the-art approaches successfully advance the frontier of depression detection through increasingly complex parameterization and integration, our work serves a distinctly different, yet highly complementary, analytical purpose. Rather than proposing another complex fusion architecture, this study aims to conceptually investigate the intrinsic limitations of integrating multiple modalities within pre-trained models. The rapid proliferation of these complex fusion networks underscores a critical need to understand the underlying mechanics of modality interaction. Our empirical analysis reveals that without careful control, the inherent feature entanglement during multimodal integration can severely degrade performance, particularly in scenarios with limited data or dominant shortcut cues. Therefore, by analyzing the failures of pre-trained models and explicitly demonstrating the benefits of information separation, our study provides a fundamental diagnostic perspective. It highlights that understanding and resolving feature entanglement is an essential prerequisite for designing the next generation of multimodal fusion systems.}

\section{Baseline System}\label{baseline system}
Our methodology, shown in Figure \ref{fig:overview}, follows a structured three-phase approach. In the first phase, we extract acoustic landmarks from speech and perform various data processing techniques. The next phase, cross-modal instruction fine-tuning, involves guiding the LLM to learn the specific characteristics and nuances of these acoustic landmarks. In the final phase, P-Tuning, the LLM is refined to apply this understanding effectively for depression diagnosis.

\subsection{Landmark Extraction}
\begin{table}[ht]
\centering
\caption{Descriptions of the six investigated landmarks}
\label{landmark_table}
\begin{tabular}{l|p{0.7\linewidth}}
\hline
\hline
\textbf{Landmark} & \textbf{Description} \\
\hline
p & Start (+) or end (–) of periodicity \\[5pt]
v & Onset (+) or offset (–) of voiced frication \\[5pt]
s & Release (+) or closure (–) of a nasal \\[5pt]
f & Onset (+) or offset (–) of frication \\[5pt]
g & Start (+) or end (–) of vocal fold vibrations \\[5pt]
b & Onset (+) or offset (–) of turbulent noise during obstruent regions \\
\hline
\end{tabular}
\end{table}

Figure \ref{fig:landmark} illustrates acoustic landmarks as a discretization of speech signals into acoustic tokens. Our study utilizes a Python-based landmark detection algorithm, which builds on the approaches of Liu~\cite{liu1996landmark} and Boyce~\cite{boyce2012speechmark}. The detection process begins by dividing the spectrogram into six frequency bands, where energy shifts are analyzed in two detection stages to identify specific landmarks. For \textbf{glottal (g)} landmarks, marking the onset and offset of vocal fold vibrations, we monitor low-frequency bands for abrupt energy increases or decreases exceeding a 5 dB threshold. To capture physiological accuracy, we ensure each glottal onset has a corresponding offset using dynamic programming, reflecting natural glottal cycles. \textbf{Burst (b)} landmarks, associated with plosive sounds, are detected through sharp energy increases in the mid-frequency bands, using an 8 dB threshold to differentiate bursts from other acoustic events. Syllabic regions, represented by \textbf{syllabic (s)} landmarks, are identified based on gradual energy buildup or drop within the mid to high-frequency bands, with a 6 dB threshold used to capture sustained energy characteristic of vowels or sonorant consonants.

Further, to detect \textbf{frication (f)} and \textbf{voiced Frication (v)} landmarks, we examine power shifts within the high-frequency bands and adjust energy levels in the low-frequency bands to differentiate fricative sounds in unvoiced and voiced regions. For detecting \textbf{periodicity (p)} landmarks, autocorrelation calculations are applied to audio frames, capturing recurring patterns indicative of voiced speech. Finally, to address data limitations, we augment the dataset by sampling sub-dialogues, balancing depression cases, and leveraging consecutive landmark pairs to efficiently represent the sequence of acoustic events in each speech segment. This method captures essential acoustic and timing information necessary for effective analysis in depression detection tasks.
\begin{figure}[t]
    \centering
    \includegraphics[width=0.5\textwidth]{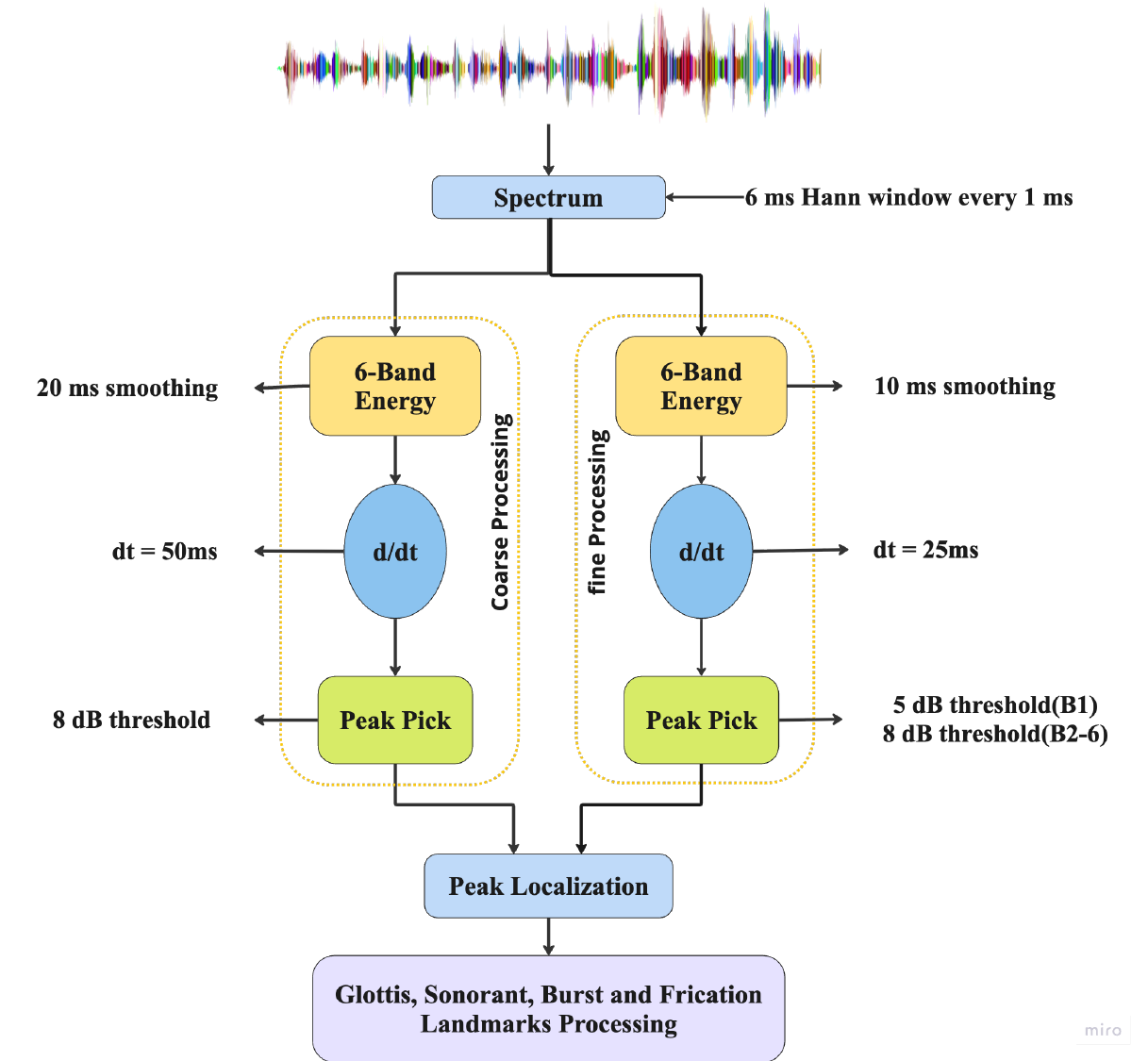}
    \caption{Landmark detection system}\vspace{-5mm}
    \label{fig:filter}
\end{figure}

\subsection{Data Augmentation}
Clinical depression assessments typically involve assigning a single label to each session conducted through interviews. This labeling, when applied to a fixed dataset, results in fewer samples compared to the abundant utterances and frames available in other speech tasks, creating challenges of data scarcity. Additionally, severe data imbalance persists, as healthy cases (positive) are significantly fewer than depression cases (negative). To address this, we adopted the sub-dialogue shuffling approach from Wu et al.~\cite{wu2023self}, sampling segments \( x_{s:e} \) from complete dialogues \( x_{1:T} \), with \( s \) and \( e \) indicating random start and end utterances.

Using acoustic landmarks enables us to extend the sub-dialogue length beyond traditional frame-based approaches while maintaining computational efficiency. We maintain dataset consistency with Wu's method by using 1,000 sub-dialogue samples (\textit{M}=1000). Building on prior findings that landmark patterns are more informative than individual landmarks~\cite{huang2019investigation}, we group consecutive landmark pairs to capture richer temporal patterns while reducing sequence length, as illustrated in Figure \ref{fig:landmark}.

\subsection{Hint Cross-modal Instruction Fine-Tuning}
Our approach begins by equipping the LLM with an understanding of acoustic landmarks, an essential step since LLMs are not natively exposed to this type of data. This foundational training enables the model to recognize, interpret, and utilize acoustic landmark information effectively.

In the central part of Figure \ref{fig:overview}, we illustrate our method, which involves instructing the LLM to predict probable acoustic landmarks based on textual input. This process serves two purposes: it familiarizes the LLM with acoustic landmarks and establishes a connection between the speech (landmarks) and text modalities through paired data. To achieve efficient adaptation, we incorporate low-rank matrices into the Query and Key matrices of the self-attention layer via LoRA~\cite{hu2022lora}. We also adjust the embedding layer to accommodate the newly integrated landmarks in the vocabulary. Throughout training, we update the \textbf{embedding layer}, \textbf{linear head}, and \textbf{LoRA matrices} to ensure the effective integration of landmark features. The objective is to minimize the negative log-likelihood across all samples, including the prefix, expressed as:
\begin{equation}
    \mathcal{L}(M|C) = - \sum_{j=1}^{x} \sum_{i=1}^{y_j} \log P(s_{i,j}|s_{<i,j},M),
\end{equation}
where \( x \) represents the dataset sample count \( C \), \( y_j \) corresponds to text with associated landmarks in each sample \( S \), and \( M \) is the fine-tuned language model.

Our cross-modal instruction fine-tuning process incorporates contextual hints about the depression status of the speakers. Experimental results show that these hints significantly improve the LLM's performance in depression detection. This improvement in performance when provided with depression-related context suggests that acoustic landmarks indeed capture diagnostically relevant patterns - an observation that aligns with prior findings about the relationship between speech patterns and depression~\cite{cummins2015review, huang2019investigation}.

\subsection{P-Tuning for Depression Detection}
In the initial phase, we trained the LLMs to comprehend the concept of landmarks. Building on this foundation, we applied P-tuning~\cite{LIU2023} to enable the LLMs to integrate both text and landmark information for effective depression detection. For this purpose, we replaced the language modeling head with a classification layer. The training objective was to minimize the cross-entropy loss for classification, represented as:
\begin{equation}
    \mathcal{L}= -\sum_{c=1}^{C} y_{o,c} \log(p_{o,c}),
\end{equation}
where \( C \) is the total number of classes, \( y_{o,c} \) serves as an indicator variable set to 1 if the observation \( o \) belongs to class \( c \), and 0 otherwise. \( p_{o,c} \) denotes the predicted probability that observation \( o \) falls within class \( c \).

Our experiments with different fine-tuning approaches revealed that directly applying LoRA across all layers of Llama2 yielded better performance than using manual templates with P-tuning. Based on these findings, we adopted the full-layer LoRA approach for our subsequent experiments with the baseline system.

\subsection{Baseline Experimental Setup}
\textbf{Dataset}. The DAIC-WOZ dataset~\cite{devault2014simsensei} is widely regarded as a benchmark for multimodal depression detection, comprising 189 recorded clinical interviews between interviewers and patients. In the training set, 30 out of 107 interviews are marked as depressed, while the development set includes 12 cases of depression among 35 interviews. Consistently with previous studies~\cite{gong2017topic,shen2022automatic,wu2022climate,wu2023self}, we evaluated our results using the development subset.
\begin{table}[t]
    \centering
    \caption{F1 scores across LLM and speech models by input type. Text-based, landmark-based, and combined inputs were evaluated on various LLM configurations, with Speech SSL models (with data augmentation) included for comparison. \textcolor{black}{Speech SSL models use CNN-based downstream classifiers as reported in the original study~\cite{wu2023self}, while our LLM-based approaches use classification heads. Additional results show the impact of data augmentation dependency: Llama2-13B without data augmentation and WavLM with reduced augmentation (100 vs 1000 samples).}}
    \setlength{\abovetopsep}{0pt}
    \setlength{\belowbottomsep}{0pt} 
    \setlength{\aboverulesep}{0pt} 
    \setlength{\belowrulesep}{0pt}
    \renewcommand{\arraystretch}{1.2} 
    \setlength{\tabcolsep}{12pt} 
    \begin{tabular}{llc}
    \toprule
    \toprule
    
    \textbf{Input Type} & \textbf{Model} & \textbf{F1 Score} \\
    \midrule
    
    \multirow{4}{*}{\textbf{Text}} 
    & Llama2-7B & 0.578 \\
    & Llama2-7B Chat & 0.488 \\
    & Llama2-13B & \textbf{0.636} \\
    & Llama2-13B Chat & 0.545 \\
    & \textcolor{black}{  Chat(w/o aug)~\cite{arcan2024assessment}} & \textcolor{black}{0.235} \\
    \midrule
    \multirow{2}{*}{\textbf{Zero Shot-Text}} 
    & GPT-3.5 & 0.545 \\
    & GPT-4 & \textbf{0.571} \\
    \midrule
    
    \multirow{4}{*}{\textbf{Landmark}} 
    & Llama2-7B & 0.521 \\
    & Llama2-7B Chat & 0.434 \\
    & Llama2-13B & \textbf{0.559} \\
    & Llama2-13B Chat & 0.538 \\
    \midrule
    
    \multirow{4}{*}{\textbf{Text + Landmark}} 
    & Llama2-7B & 0.545 \\
    & Llama2-7B Chat & 0.500 \\
    & Llama2-13B & \textbf{0.695} \\
    & Llama2-13B Chat & 0.666 \\
    & \textcolor{black}{Llama2-13B(w/o aug)} & \textcolor{black}{0.33} \\
    \midrule
    
    \multirow{3}{*}{\textbf{Speech SSL Models} \cite{wu2023self}} 
    & Wav2vec 2.0 & 0.627 \\
    & HuBERT & 0.667 \\
    & WavLM & 0.700 \\
    & \textcolor{black}{WavLM(reduced aug)} &\textcolor{black}{0.451} \\
    \toprule
    \end{tabular}
    
    \label{Main Result}
\end{table}
\textbf{Model Configurations}.
Our experiments employed the Llama2-7B, Llama-7B Chat, Llama2-13B, and Llama2-13B Chat models~\cite{touvron2023llama}, conducted on a system with 8 NVIDIA A100 80GB GPUs. The Llama 2-Chat models were tailored for interactive, conversational tasks. During the cross-modal instruction fine-tuning stage, we fine-tuned the model for 10 epochs with a batch size of 128, LoRA rank of 8, 100 warmup steps, and a learning rate of 1e-6. For the depression detection phase, we fine-tuned the model over 8 epochs with a batch size of 256, 30 virtual tokens, 256 encoder hidden units, and a learning rate of 1e-6. In both phases, AdamW served as the optimizer, and model parallelism was implemented to facilitate efficient fine-tuning. In the ablation study, we applied hyperparameter tuning using the Tree-structured Parzen Estimator (TPE) method~\cite{bergstra_algorithms_2011}.

\subsection{Baseline Comparison with Previous Pre-trained Model Results}

Table \ref{Main Result} summarizes the F1 scores achieved by Llama2 models across various configurations for depression detection, while also comparing these results with those of GPT-3.5 and GPT-4 in the text-only modality. Notably, GPT-3.5 and GPT-4 were not fine-tuned specifically for this task; rather, we utilized meticulously designed prompts to guide the models in evaluating whether each sample originated from a patient with depression.

For the ‘landmark’ and ‘text + landmark’ modalities, our process first entailed cross-modal instruction fine-tuning with hints, followed by P-tuning to optimize for depression detection. This approach aimed to equip the LLMs with a foundational understanding of landmarks before advancing to the diagnostic stage.

The findings show that LLMs relying solely on text for depression detection yield relatively modest performance across all models, even including state-of-the-art models like GPT-3.5 and GPT-4, which typically excel in various applications. This performance shortfall can be attributed to two primary factors. Firstly, \textbf{text alone is limited in conveying emotional context}, as it lacks the nuances found in speech. For example, the phrase “It’s raining today” might evoke a positive sentiment for some while a negative sentiment for others. Although text alone leaves these interpretations ambiguous, speech data could capture the emotional tone more accurately. Secondly, \textbf{data constraints also impact model performance}. Labels are assigned only at the document level, and there are limited data samples available currently, since there is no large public dataset for multimodal depression detection. These factors, combined with the lack of data granularity, restrict the model's capacity for accurate detection.

Integrating landmarks improved the performance across all models, validating the benefit of incorporating landmark-based information. Landmarks convey affective variations, embedding additional acoustic information that aids LLMs in identifying depression. However, relying solely on landmarks remained suboptimal for depression detection, likely due to the fact that even after cross-modal fine-tuning, integrating information exclusively from other modalities (like audio or visual) can reduce the stability of LLMs~\cite{zhang2023llama,li2023prompting}.

\section{Current Limitations and Discussion}\label{Limitation}
\begin{table}[t]
\centering
\setlength{\abovetopsep}{0pt}
\setlength{\belowbottomsep}{0pt} 
\setlength{\aboverulesep}{0pt} 
\setlength{\belowrulesep}{0pt}
\renewcommand{\arraystretch}{1.2}
\caption{Average and maximum F1 scores for Wav2Vec, WavLM, and HuBERT at specific layers with SVM classifier \textcolor{black}{without any data augmentation}}
\resizebox{\columnwidth}{!}{%
\begin{tabular}{l|c|c|c}
\toprule
\toprule
\textbf{Model} & \textbf{Layer} & \textbf{Average F1} & \textbf{Max F1} \\
\midrule
Wav2Vec 2.0 & 8 & 0.597525 & 0.625 \\
WavLM & 8 &0.6471  & 0.6471\\
HuBERT & 10 & 0.5707  &  0.6286 \\
\toprule
\end{tabular}%
}
\label{tab:ssl_F1_Scores}
\end{table}
\subsection{Limitations of Pre-trained Models in Depression Detection}

Existing methods for depression detection using LLMs and speech SSL models generally rely on data augmentation to increase dataset size, as shown in Table~\ref{Main Result}. \textcolor{black}{While data augmentation is a common and valid technique in machine learning, our concern relates to the extent of reliance on data augmentation required for SSL models to achieve reasonable performance in depression detection. For instance, previous studies have shown that when data augmentation is reduced from 1000 random sub-dialogue samples to 100 samples, WavLM's performance drops dramatically from 0.700 to 0.451, indicating heavy dependence on extensive data augmentation~\cite{wu2023self}.} This approach contradicts the foundational aim of SSL models, which are designed to generalize across a variety of downstream tasks with minimal labeled data. Table~\ref{tab:ssl_F1_Scores} shows the performance of speech SSL models on depression detection without data augmentation. We selected the best-performing layer based on previous studies~\cite{wu2023self} and conducted experiments on the development set to ensure fair comparisons and thorough analysis. Using random hyperparameter search, we report both the average and maximum F1 scores achieved, applying the same methodology in subsequent experiments.

As seen in Table~\ref{tab:ssl_F1_Scores}, the absence of data augmentation results in a notable decline in performance for speech SSL models in the depression detection task. Additionally, information in speech SSL models is distributed across multiple layers~\cite{chen2022wavlm,hsu2021hubert}, highlighting the limitations of relying solely on a single layer for this purpose.

Shifting the focus to LLMs, their application to depression detection presents a distinct set of challenges. Specifically, LLMs also require data augmentation to expand dataset size to perform depression detection, which adds complexity to the training process. \textcolor{black}{As shown in Table~\ref{Main Result}, LLM-based approaches exhibit catastrophic performance degradation when data augmentation is removed (e.g., Llama2-13B Chat dropping from 0.666 to 0.235). This extensive dependence poses particular challenges as training LLMs with such extensive task-specific data augmentation may compromise their general-purpose capabilities~\cite{kothaunderstanding,huang2024mitigating}, causing models to become overly specialized for specific augmented data distributions rather than maintaining their broad knowledge and reasoning abilities.} Moreover, integrating new modalities into LLMs compromises their inherent performance~\cite{zhang2023llama,li2023prompting}, underscoring the need for optimized approaches in multimodal depression detection.

\subsection{Information Entanglement Hypothesis}
\begin{figure}[ht]
    \centering
    \begin{subcaptionbox}{Decision boundary for two independent data\label{fig:fig1}}{
        \includegraphics[width=0.45\columnwidth]{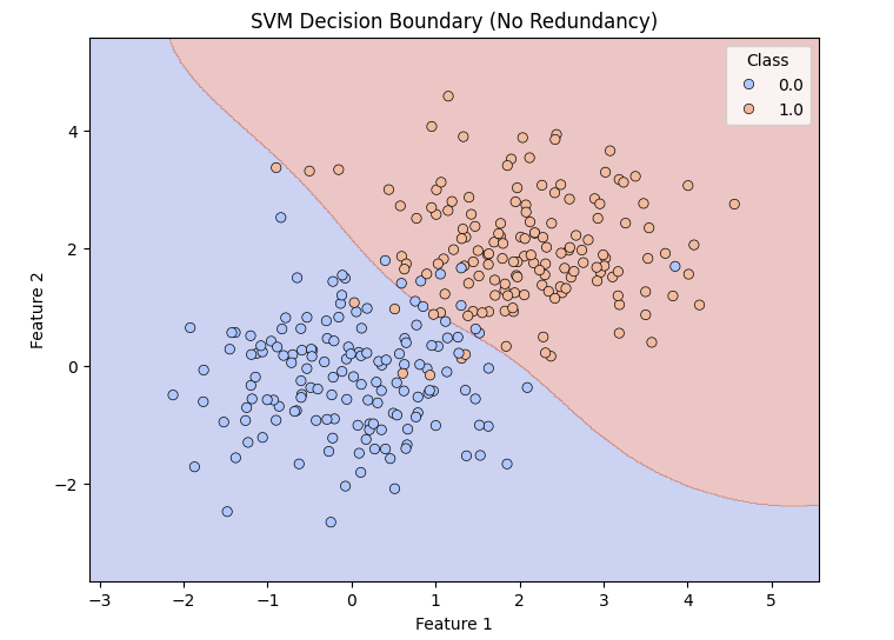}}
    \end{subcaptionbox}
    \hfill
    \begin{subcaptionbox}{The decision boundary for the data contains overlapping information\label{fig:fig2}}{
        \includegraphics[width=0.45\columnwidth]{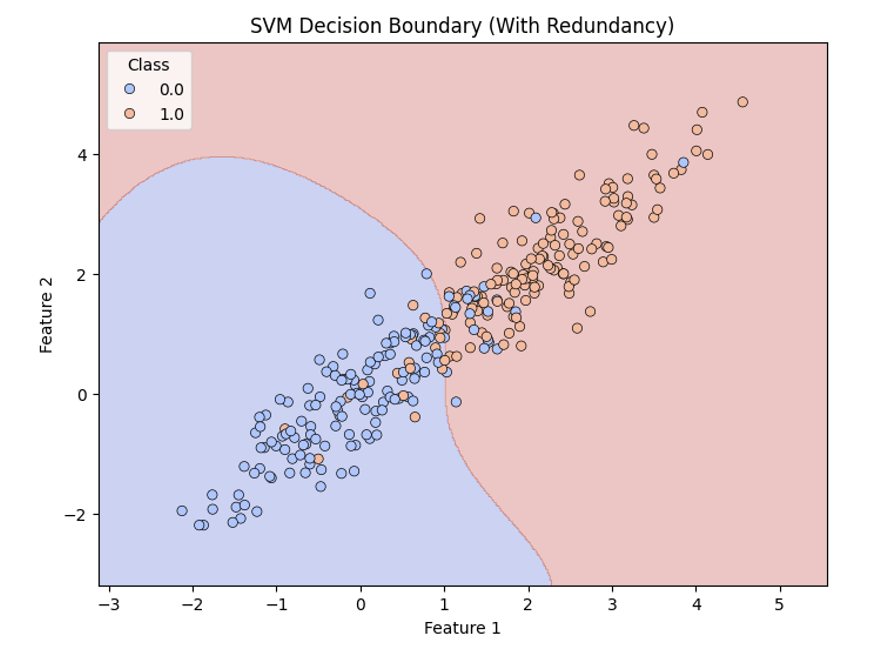}}
    \end{subcaptionbox}
    \caption{A comparison of decision boundary for dependent/independent Data}
    \label{fig:side_by_side}
\end{figure}

\textcolor{black}{The empirical limitations identified in Section IV.A - including SSL models' heavy dependence on data augmentation and LLMs' performance degradation when integrating new modalities - point to a fundamental representational issue rather than simple capacity constraints. To explain these phenomena, we propose the information entanglement hypothesis as the underlying cause of these observed limitations.}

Based on our analysis of pre-trained models' limitations, we formulate a specific hypothesis about their poor performance in depression detection. Prior work has shown that both content and speech features contain depression-relevant information~\cite{zheng2023two}. However, we hypothesize that the simultaneous encoding of these features within SSL model representations impairs the model's ability to learn effective decision boundaries for depression detection. 


To validate this hypothesis, we first demonstrate the impact of feature entanglement through controlled experiments with simulated data. \textcolor{black}{We generated two datasets with 1000 samples each: an independent features dataset where two features (X1, X2) are independently sampled from normal distributions with distinct class centers, and an entangled features dataset where X2 = X1 + noise, creating correlation that simulates the mixing of content and speech information in SSL representations. We evaluated classification performance using SVM with RBF kernel and measured mutual information between features and labels to quantify information relationships.} Figure~\ref{fig:side_by_side} presents two scenarios of binary classification: one with independent features (Figure~\ref{fig:fig1}) and another with entangled features (Figure~\ref{fig:fig2}). When features are independent, the classifier establishes a clear, linear decision boundary \textcolor{black}{with superior performance (F1-score: 0.944, Accuracy: 0.943)}. In contrast, entangled features lead to a complex, non-linear decision boundary with regions of high uncertainty \textcolor{black}{and degraded performance (F1-score: 0.875, Accuracy: 0.873)}. \textcolor{black}{Quantitatively, our simulation demonstrates that feature entanglement results in reduced mutual information for the correlated feature (0.301 vs 0.336 for independent features) and an overall performance drop of approximately 7\%, validating our hypothesis that feature mixing creates substantial classification challenges.} This simulated demonstration aligns with our hypothesis that feature entanglement in SSL models creates similar classification challenges in depression detection.

To empirically test this hypothesis, we develop an information separation framework that explicitly disentangles content and speech features within SSL representations. If our hypothesis holds, this separation should lead to improved classification performance compared to standard SSL approaches that allow feature entanglement. \textcolor{black}{Moreover, this approach should address both limitations identified in Section IV.A: eliminating the need for extensive data augmentation in SSL models and enabling effective multimodal integration in LLMs without performance degradation.}

\begin{figure*}[t]
    \centering
    \includegraphics[width=1\textwidth]{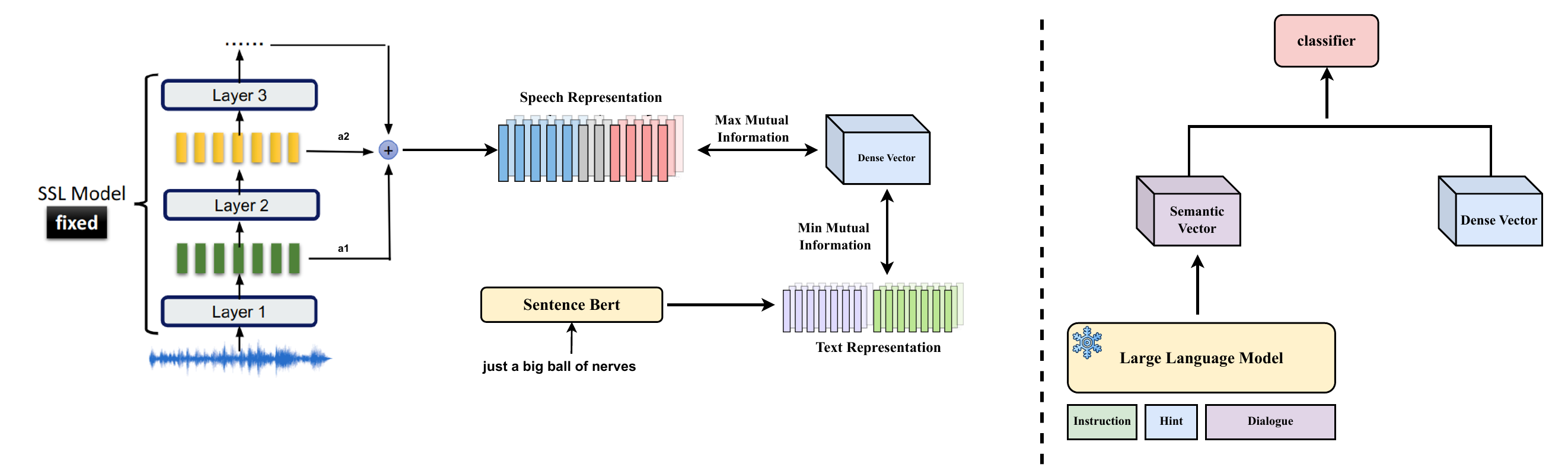}
    \caption{Overview of the proposed information separation system for depression detection. On the left, the system separates information, constructing dense vectors that contain only speech-specific information. On the right, these dense vectors are integrated with an LLM for performing depression detection.} \vspace{-5mm}
    \label{fig:pipeline}
\end{figure*}

\section{Information Separation Framework for Depression Detection}\label{new system}

Figure~\ref{fig:pipeline} introduces our proposed framework for addressing the limitations of pre-trained models in depression detection. Motivated by our hypothesis about feature entanglement, we design an information separation module (left side) that explicitly disentangles modality-specific features into dense vectors. On the right side, we demonstrate how these separated features can be effectively integrated with frozen LLM embeddings for depression detection.

Through this framework, we investigate three research questions that directly address the limitations identified in pre-trained models:

\emph{RQ1:} Can explicit separation of speech-specific and content-specific features improve depression detection performance? This question tests our core hypothesis about feature entanglement.

\emph{RQ2:} How does utilizing information across multiple SSL model layers impact depression detection performance? This explores whether the limitation stems from single-layer feature extraction.

\emph{RQ3:} Can frozen LLM embeddings with appropriate prompting achieve better performance than fine-tuning approaches? Here we examine whether fine-tuning might actually impair the model's ability to leverage its pre-trained knowledge.

Each research question motivates a series of experiments in our subsequent analysis, designed to systematically evaluate our proposed solution against the limitations of current approaches.

\subsection{Design of the Information Separation System}

Building on our hypothesis about feature entanglement, we first design and evaluate systems for explicit information separation. We develop two parallel approaches: one that isolates speech-specific information and another that preserves only text-specific information. This design allows us to systematically test whether separated features outperform entangled representations in depression detection.

To achieve this separation, we implement a contrastive learning framework~\cite{gao2021simcse} that explicitly disentangles modality-specific information from the mixed representations in SSL models. The framework uses contrastive objectives to maximize the preservation of one modality's information while minimizing interference from the other, resulting in specialized dense vectors for each modality. This approach directly tests our hypothesis about the detrimental effects of feature entanglement in pre-trained models.

The process begins by aggregating information distributed across the layers of a speech SSL model. Let $\mathbf{H} = \{h_1, h_2, \ldots, h_L\}$ represent the hidden states from $L$ layers for a given input. We compute a weighted sum~\cite{yang2021superb} of these hidden states to form a compact representation:
\begin{equation}
    h_{\text{speech}}^{(i)} = \sum_{l=1}^{L} \alpha_i^{(l)} h_l^{(i)}
\end{equation}
where $h_l^{(i)}$ is the hidden state of the $l$-th layer for sample $i$, and $\alpha_i^{(l)}$ are learnable weights normalized such that $\sum_{l=1}^{L} \alpha_i^{(l)} = 1$. This weighted sum captures relevant information across layers for downstream tasks.

Next, we project the aggregated speech features $h_{\text{speech}}^{(i)}$ and the corresponding sentence-level features $s^{(i)}$ into a shared latent space using dense networks:
\begin{equation}
    z_{\text{speech}}^{(i)} = \text{LeakyReLU}(\mathbf{W}_{\text{speech}} h_{\text{speech}}^{(i)} + \mathbf{b}_{\text{speech}})
\end{equation}
\begin{equation}
    z_{\text{sentence}}^{(i)} = \text{LeakyReLU}(\mathbf{W}_{\text{sentence}} s^{(i)} + \mathbf{b}_{\text{sentence}})
\end{equation}
where $\mathbf{W}_{\text{speech}}$ and $\mathbf{W}_{\text{sentence}}$ are learnable weight matrices, and $\mathbf{b}_{\text{speech}}$ and $\mathbf{b}_{\text{sentence}}$ are biases. These projections ensure that both modalities are represented in a comparable space.

The dense vector $\mathbf{v} \in \mathbb{R}^d$ is trained to retain speech-specific information. This is achieved by maximizing the information between $\mathbf{v}$ and $z_{\text{speech}}$ while minimizing its information with $z_{\text{sentence}}$. Information is measured using cosine similarity:
\begin{equation}
    \text{sim}_{\text{speech}} = \frac{\mathbf{v} \cdot z_{\text{speech}}}{\|\mathbf{v}\| \|z_{\text{speech}}\|}
\end{equation}
\begin{equation}
    \text{sim}_{\text{sentence}} = \frac{\mathbf{v} \cdot z_{\text{sentence}}}{\|\mathbf{v}\| \|z_{\text{sentence}}\|}
\end{equation}

To achieve this objective, we employ a contrastive loss function:
\footnotesize
\begin{equation}
    \mathcal{L}_{\text{Loss}} = -\frac{1}{N} \sum_{i=1}^N \log \frac{\exp(\text{sim}_{\text{speech}}^{(i)} / \tau)}{\exp(\text{sim}_{\text{speech}}^{(i)} / \tau) + \exp(\text{sim}_{\text{sentence}}^{(i)} / \tau)}
\end{equation}
\normalsize
where $\tau$ is a temperature parameter that controls the sharpness of the similarity distribution. This loss ensures that the dense vector aligns closely with the speech features while diverging from text-derived information, effectively filtering out content not unique to speech.

By optimizing this loss, the dense vector $\mathbf{v}$ learns to isolate speech-specific information. Parameters updated during training include the dense network weights ($\mathbf{W}_{\text{speech}}$, $\mathbf{W}_{\text{sentence}}$), the dense vector $\mathbf{v}$, and the layer weights $\alpha_i^{(l)}$. 

When the goal is to enable the dense vector to capture only content-related information, we adjust the contrastive loss function accordingly. Instead of emphasizing the separation of speech-specific information, the modified loss function promotes similarity between the dense vector and sentence embeddings while minimizing its similarity with speech embeddings. The reformulated contrastive loss is defined as follows:

\footnotesize
\begin{equation}
\mathcal{L}_{\text{Loss}} = -\frac{1}{N} \sum_{i=1}^N \log \frac{\exp(\text{sim}_{\text{sentence}}^{(i)} / \tau)}{\exp(\text{sim}_{\text{sentence}}^{(i)} / \tau) + \exp(\text{sim}_{\text{speech}}^{(i)} / \tau)}
\end{equation}
\normalsize

where \( \text{sim}_{\text{sentence}}^{(i)} \) and \( \text{sim}_{\text{speech}}^{(i)} \) represent the cosine similarity of the dense vector with the sentence and speech projections, respectively, for the \( i \)-th sample. By focusing on sentence similarity, this loss function encourages the dense vector to align more closely with content-related information.

To validate the effectiveness of multi-layer SSL features in depression detection and respond to research question two, we designed an experiment that solely maximizes the similarity between the dense vector and the weighted sum of SSL features. The reformulated loss is defined as follows:
\begin{equation}
\mathcal{L}_{\text{loss}} = -\frac{\mathbf{v} \cdot z_{\text{speech}}}{\|\mathbf{v}\| \|z_{\text{speech}}\|}
\end{equation}

\subsection{Integration of Separated Features with LLMs}
The second component of our framework addresses how to effectively combine separated features with LLM capabilities (right side of Figure~\ref{fig:pipeline}). Rather than fine-tuning the LLM, we keep it frozen and use explicit task-specific prompts ("hints") to guide the extraction of relevant semantic embeddings. This design choice stems from our observation that fine-tuning might disrupt the model's pre-trained knowledge. For example, we use prompts like:

\begin{quote}
\textit{“Extract the semantic embedding from the following dialogue \textbf{for depression detection}.”}
\end{quote}

This multimodal framework combines the acoustic features extracted from speech SSL models with semantic information derived from LLMs, providing a comprehensive representation for depression detection. The speech dense vectors, denoted as $\mathbf{v}_{\text{speech}}$, are constructed by aggregating sentence-level dense vectors generated by the SSL model. Specifically, the dense vector for each sample is computed by taking the mean of all sentence-level dense vectors:
\begin{equation}
\mathbf{v}_{\text{speech}} = \frac{1}{N} \sum_{i=1}^N \mathbf{v}_{\text{speech}}^{(i)},
\end{equation}
where $\mathbf{v}_{\text{speech}}^{(i)}$ represents the dense vector for the $i$-th sentence, and $N$ is the total number of sentences in the dialogue. This aggregation captures both global and local acoustic information, which is crucial for identifying depression-related patterns.

Semantic embeddings were extracted using a pre-trained LLM such as LLaMA. To ensure the embeddings were task-specific, the LLM processes each dialogue using the tailored hint described above. The dialogue was encoded, and the hidden states of the final layer were mean-pooled to generate a fixed-size semantic embedding:
\begin{equation}
\mathbf{E}_{\text{LLM}} = \frac{1}{T} \sum_{t=1}^T \mathbf{h}_{\text{LLM}}^{(t)},
\end{equation}
where $\mathbf{h}_{\text{LLM}}^{(t)}$ denotes the hidden state at time step $t$, and $T$ represents the sequence length. This embedding captures the semantic content of the dialogue in alignment with the depression detection task.

To improve computational efficiency and reduce redundancy in the high-dimensional LLM embeddings, we applied dimensionality reduction using UMAP~\cite{mcinnes2018umap}:
\begin{equation}
\mathbf{E}_{\text{LLM, reduced}} = \text{UMAP}(\mathbf{E}_{\text{LLM}}),
\end{equation}
where we reduce the embedding dimension to 300 based on preliminary experiments that showed this preserves task-relevant information while reducing computational overhead.

The final multimodal representation combines the speech dense vector $\mathbf{v}_{\text{speech}}$ and the reduced LLM embedding $\mathbf{E}_{\text{LLM, reduced}}$:
\begin{equation}
\mathbf{F} = [\mathbf{v}_{\text{speech}}; \mathbf{E}_{\text{LLM, reduced}}],
\end{equation}
where $[\cdot; \cdot]$ denotes concatenation. This integration creates a unified feature vector that preserves the separated modality-specific information while allowing their complementary use.

For classification, we deliberately choose a simple SVM classifier to ensure that any performance improvements can be attributed to our feature separation approach rather than classifier complexity. This allows us to directly evaluate how the quality of separated features impacts depression detection performance. To validate our framework's effectiveness, we conduct experiments using both combined ($\mathbf{F}$) and independent modality features ($\mathbf{v}_{\text{speech}}$ and $\mathbf{E}_{\text{LLM, reduced}}$), focusing on how feature separation affects model performance.

\section{Evaluating the Information Separation Hypothesis}\label{result}
\subsection{Experimental Setup for Information Separation}
To systematically evaluate our hypothesis about feature entanglement, we conduct experiments across three scenarios using major SSL models (HuBERT~\cite{hsu2021hubert}, Wav2Vec2~\cite{baevski2020wav2vec}, and WavLM~\cite{chen2022wavlm}):

1. Speech-specific information preservation

2. Text-specific information preservation 

3. Layer-wise information combination through learnable weights (ranging from 0 to 1, summing to 1 across layers)

Implementation details: batch size 64, 1500 epochs, temperature 0.1, learning rate 0.0001, Adam optimizer on V100 GPUs. Dense vector dimensions varied from 100 to 500 to study dimensionality impact.

\subsection{Validating the Benefits of Information Separation}
Our experiments with SSL models demonstrate the impact of information separation through two key findings. First, combining information across layers yields model-specific benefits - substantial improvements for Wav2Vec and HuBERT but minimal gains for WavLM (Tables~\ref{tab:dense_vectors},~\ref{tab:ssl_F1_Scores}). This suggests that the distribution of depression-relevant information varies across model architectures. More importantly, explicit separation of speech-specific and text-specific information consistently improves performance in all approaches. The speech-specific system particularly excels, outperforming even data-augmented baselines (Table~\ref{Main Result}) and achieving state-of-the-art results for SSL models in depression detection. This superior performance without data augmentation validates our core hypothesis: feature entanglement, rather than model capacity, has been the key factor limiting SSL model performance in depression detection.

\begin{table}[t] 
\centering
\scriptsize
\def\arraystretch{1.25}
\setlength{\tabcolsep}{1.5pt}
\caption{F1 Scores across dense vector dimensions for different objectives and speech SSL Models. “Speech Information Preservation” denotes dense vectors containing only speech-related information, “Text Information Preservation” focuses on text-related information, and “Weighted Sum Similarity Maximization” maximizes similarity between dense vectors and weighted Speech SSL features.}
\label{tab:dense_vectors}
\resizebox{\columnwidth}{!}{
\begin{tabular}{lccc ccc ccc}
\toprule[1.25pt]
\textbf{Dimension} & \multicolumn{3}{c}{\textbf{WavLM}} & \multicolumn{3}{c}{\textbf{Wav2Vec}} & \multicolumn{3}{c}{\textbf{HuBERT}} \\
\cmidrule(lr){2-4} \cmidrule(lr){5-7} \cmidrule(lr){8-10} 
& \textbf{F1-avg} & \textbf{F1-max} & \textbf{F1-std} 
& \textbf{F1-avg} & \textbf{F1-max} & \textbf{F1-std} 
& \textbf{F1-avg} & \textbf{F1-max} & \textbf{F1-std} \\ 
\midrule
\multicolumn{10}{c}{\textbf{Speech Information Preservation}} \\
100 & 0.552 & 0.571 & 0.014 & 0.540 & 0.546 & 0.012 & 0.549 & 0.558 & 0.006 \\
200 & 0.618 & 0.621 & 0.003 & 0.512 & 0.514 & 0.002 & 0.591 & 0.621 & 0.020 \\
300 & \textbf{0.755} & \textbf{0.769} & 0.016 & 0.525 & 0.533 & 0.010 & 0.600 & 0.615 & 0.017 \\
400 & 0.563 & 0.585 & 0.015 & 0.586 & 0.615 & 0.035 & 0.534 & 0.537 & 0.002 \\
500 & 0.634 & 0.640 & 0.012 & \textbf{0.616} & \textbf{0.640} & 0.030 & \textbf{0.676} & \textbf{0.688} & 0.022 \\
\midrule
\multicolumn{10}{c}{\textbf{Text Information Preservation}} \\
100 & 0.521 & 0.522 & 0.000 & 0.526 & 0.529 & 0.004 & 0.602 & 0.625 & 0.022 \\
200 & 0.554 & 0.564 & 0.007 & 0.602 & 0.611 & 0.012 & 0.558 & 0.558 & 0.000 \\
300 & \textbf{0.641} & \textbf{0.667} & 0.030 & 0.528 & 0.546 & 0.012 & 0.580 & 0.606 & 0.017 \\
400 & 0.608 & 0.615 & 0.015 & \textbf{0.616} & \textbf{0.621} & 0.010 & 0.530 & 0.533 & 0.006 \\
500 & 0.595 & 0.600 & 0.011 & 0.571 & 0.571 & 0.000 & \textbf{0.656} & \textbf{0.667} & 0.012 \\
\midrule
\multicolumn{10}{c}{\textbf{SSL Weighted Sum Similarity Maximization}} \\
100 & 0.563 & 0.585 & 0.017 & 0.601 & 0.625 & 0.019 & 0.552 & 0.556 & 0.008 \\
200 & \textbf{0.638} & \textbf{0.667} & 0.058 & 0.537 & 0.546 & 0.011 & 0.628 & 0.643 & 0.028 \\
300 & 0.519 & 0.522 & 0.006 & 0.550 & 0.564 & 0.020 & 0.522 & 0.522 & 0.000 \\
400 & 0.574 & 0.621 & 0.033 & 0.612 & 0.615 & 0.008 & 0.548 & 0.556 & 0.015 \\
500 & 0.557 & 0.571 & 0.010 & \textbf{0.619} & \textbf{0.632} & 0.008 & \textbf{0.627} & \textbf{0.643} & 0.031 \\
\bottomrule[1.25pt]

\end{tabular}
}
\end{table}

\textcolor{black}{\subsection{Decision Boundary Visualization and External Validation}}

\textcolor{black}{To provide visual evidence for our information entanglement hypothesis, we conducted 3D decision boundary analysis comparing original SSL features with our information-disentangled features. We trained SVM classifiers with optimized hyperparameters on both feature types and applied 3D PCA dimensionality reduction to project the high-dimensional feature space into 3D for visualization while preserving decision boundaries.}

\begin{figure}[t]
    \centering
    \begin{subcaptionbox}{\label{fig:fig5}}{\includegraphics[width=0.24\textwidth]{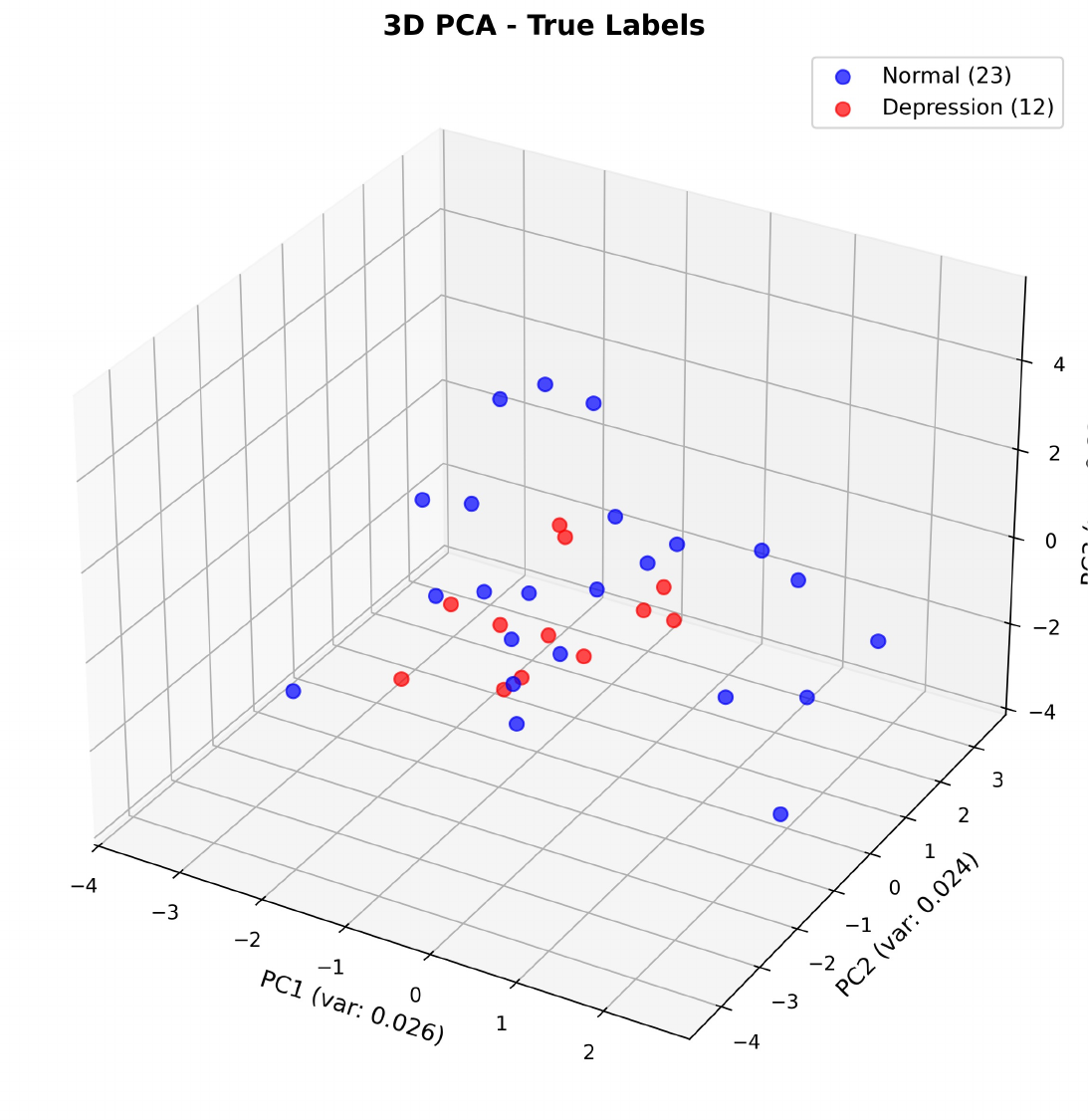}}\end{subcaptionbox}
    \begin{subcaptionbox}{\label{fig:fig6}}{\includegraphics[width=0.24\textwidth]{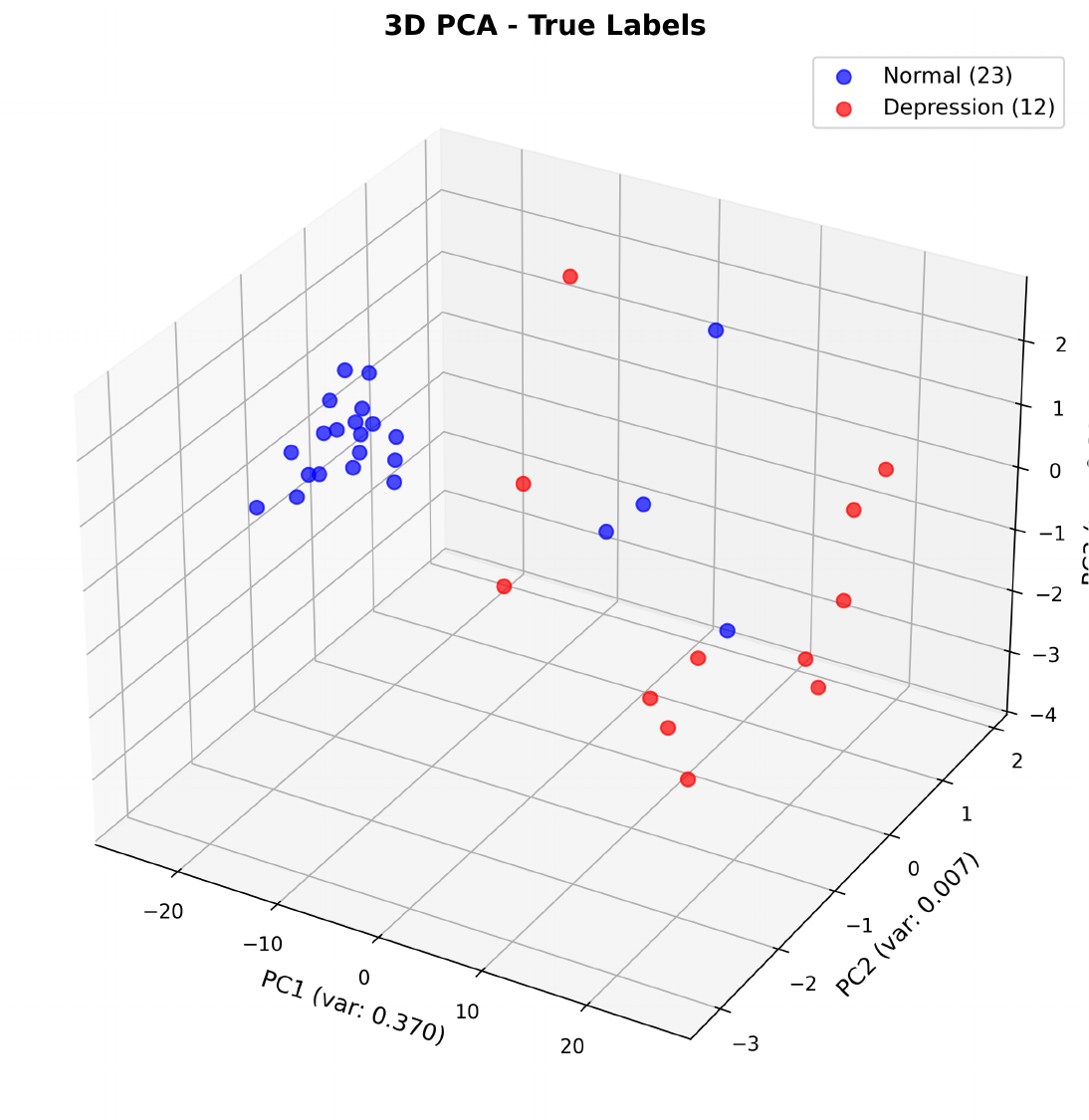}}\end{subcaptionbox}
    \caption{\textcolor{black}{Decision boundary visualization for WavLM features before and after decoupling acoustic and semantic information. Left: Original entangled features showing overlapping decision boundaries. Right: Disentangled features demonstrating clearer class separation with more distinct decision boundaries.}}
    \vspace{-14pt}
    \label{fig:real_decision_boundary}
\end{figure}

\textcolor{black}{Figure~\ref{fig:real_decision_boundary} shows the comparison between original SSL features (left) and our information-disentanglement features (right). In the left panel, data points from both classes (Normal=blue, Depression=red) show significant overlap with unclear separation, indicating confused decision boundaries. In contrast, the right panel demonstrates that after feature separation, the two classes achieve much clearer clustering with more distinct decision boundaries. This visualization directly validates our information entanglement hypothesis by showing that separated features enable more effective classification boundaries.}

\textcolor{black}{Our findings are further corroborated by recent work~\cite{10872825} that focuses on model design for depression detection. Rather than analyzing SSL model representations as we do, they operated directly on speech signals to separate specific acoustic characteristics including voiceprint, emotion, pause patterns, energy, and tremor. By explicitly separating these acoustic features, they achieved improved model performance. This work provides additional validation of our analysis from a complementary perspective, demonstrating that infomration separation benefits depression detection regardless of whether the separation occurs at the signal level or the representation level.}

\subsection{Quantifying Information Separation in Dense Vectors}

\begin{figure}[t]
    \centering
    \begin{subcaptionbox}{\label{fig:fig3}}{\includegraphics[width=0.24\textwidth]{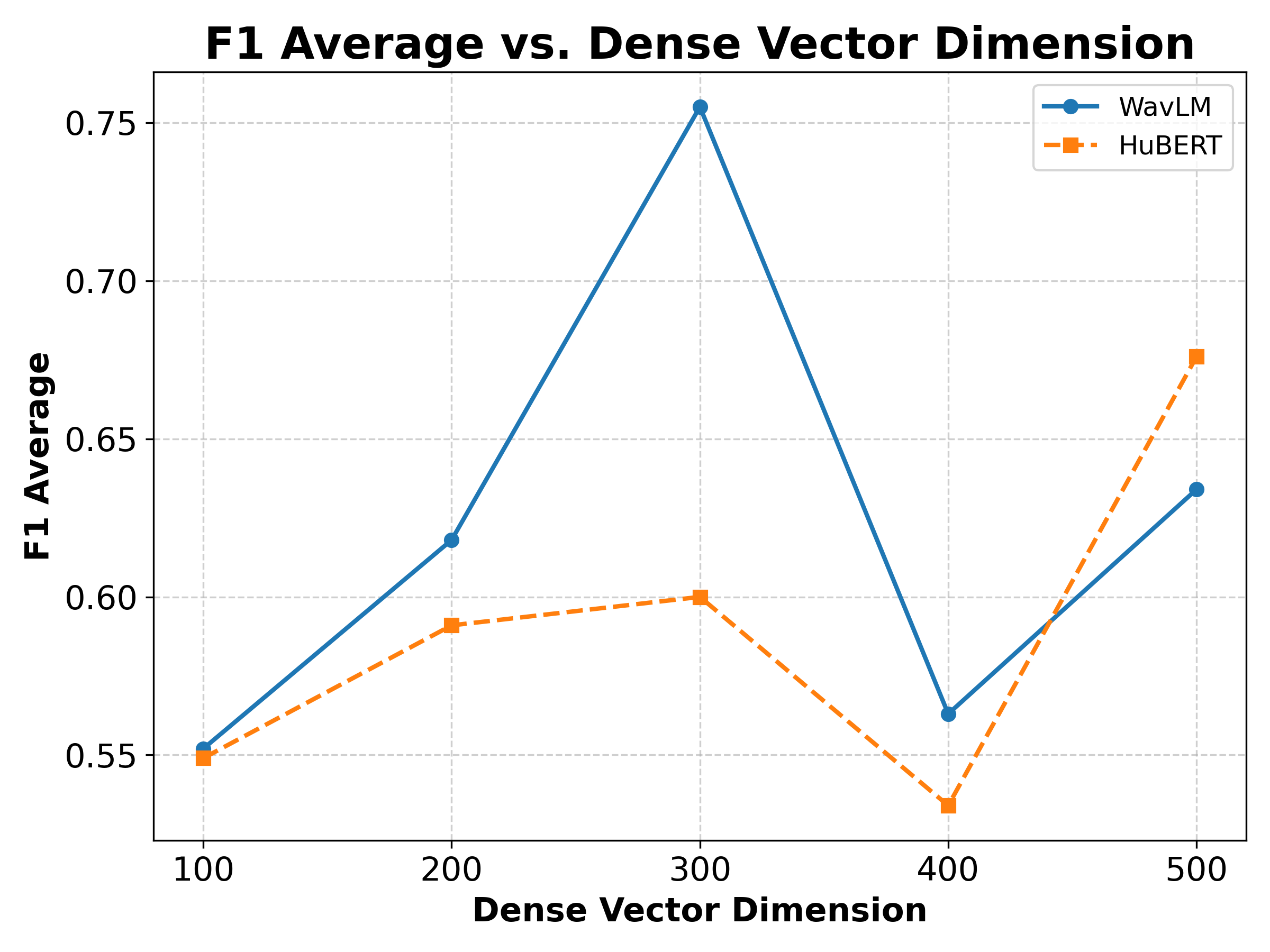}}\end{subcaptionbox}
    \begin{subcaptionbox}{\label{fig:fig4}}{\includegraphics[width=0.24\textwidth]{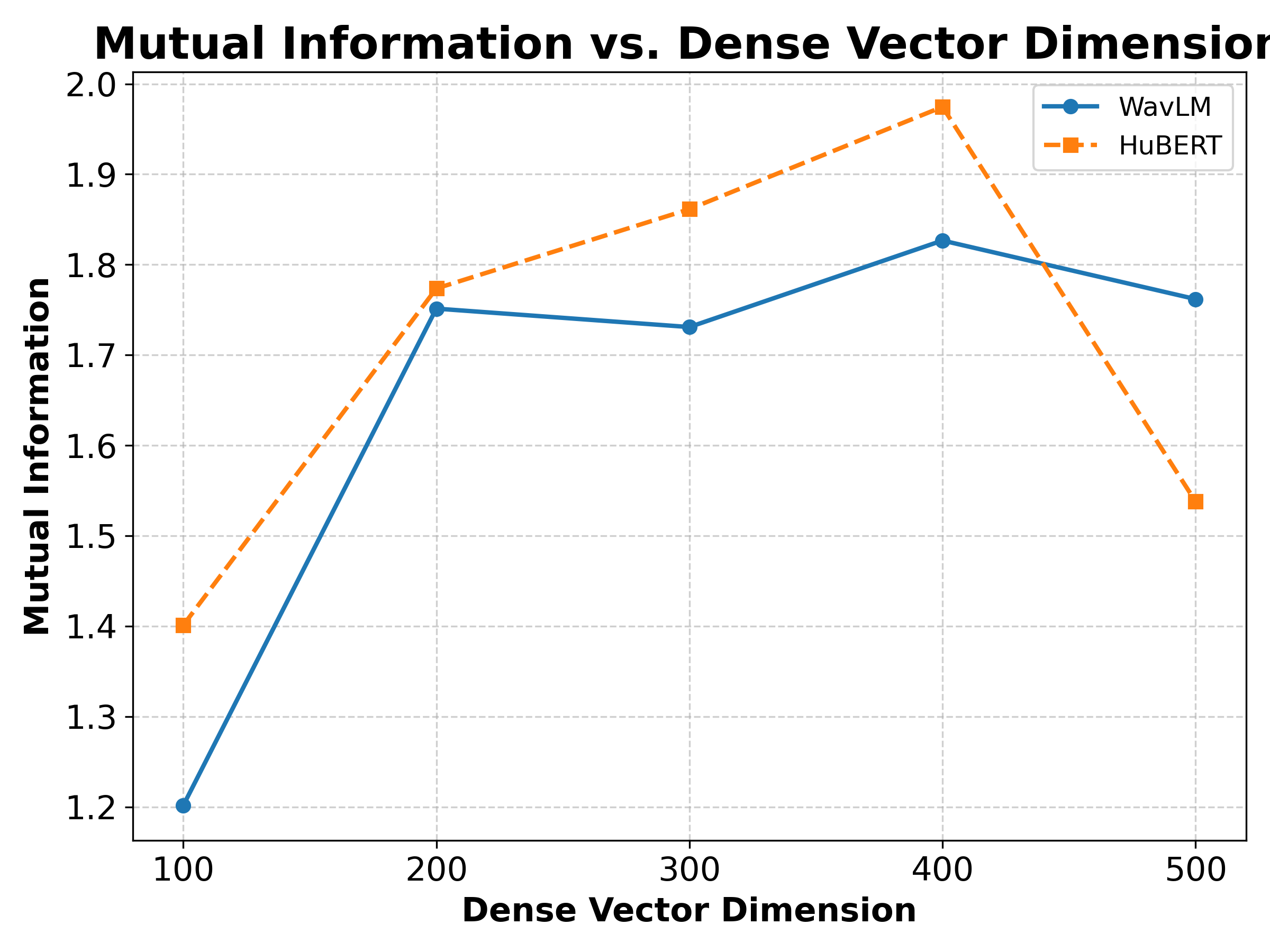}}\end{subcaptionbox}
    \caption{A comparison of mutual information and performance in Speech Dense Vector for WavLM and HuBERT}
    \vspace{-14pt}
    \label{fig:mutual_information}
\end{figure}

Building on our hypothesis that feature entanglement impairs model performance, we analyze what makes our separation approach effective. A key question emerges from the results in Figure~\ref{fig:fig3}: why do different models require different dense vector dimensions for optimal performance (300 for WavLM, 500 for HuBERT)? We propose that the effectiveness of separation depends on finding the right balance between representational capacity and information disentanglement. To verify this, we measure the degree of separation using Mutual Information Neural Estimation (MINE)~\cite{belghazi2018mine}:

\footnotesize
\begin{equation}
I_{\Theta}(\mathbf{v}, \mathbf{e}) = \sup_{\theta \in \Theta} \mathbb{E}_{P(\mathbf{v}, \mathbf{e})}[T_{\theta}(\mathbf{v}, \mathbf{e})] - \log \mathbb{E}_{P(\mathbf{v})P(\mathbf{e})}[e^{T_{\theta}(\mathbf{v}, \mathbf{e})}],
\end{equation}
\normalsize
where \(T_{\theta}\) is a trainable neural network parameterized by \(\theta\), \(P(\mathbf{v}, \mathbf{e})\) represents the joint distribution, and \(P(\mathbf{v})P(\mathbf{e})\) is the product of the marginal distributions.Our analysis reveals that performance peaks when mutual information reaches its minimum (Figure~\ref{fig:fig4}), indicating that better separation directly leads to better depression detection. However, achieving this separation requires sufficient dimensionality - too few dimensions (as seen at 100) constrain the model's ability to properly separate features while preserving task-relevant information~\cite{belghazi2018mine,geiger2021information}. This explains why optimal dimensions vary across models: each architecture requires different representational capacity to achieve effective separation.

\subsection{Layer-Level Analysis of Information in Dense Vector}
\begin{figure*}[!t]
    \centering
    \begin{subfigure}[b]{0.24\textwidth}
        \centering
        \includegraphics[width=\linewidth]{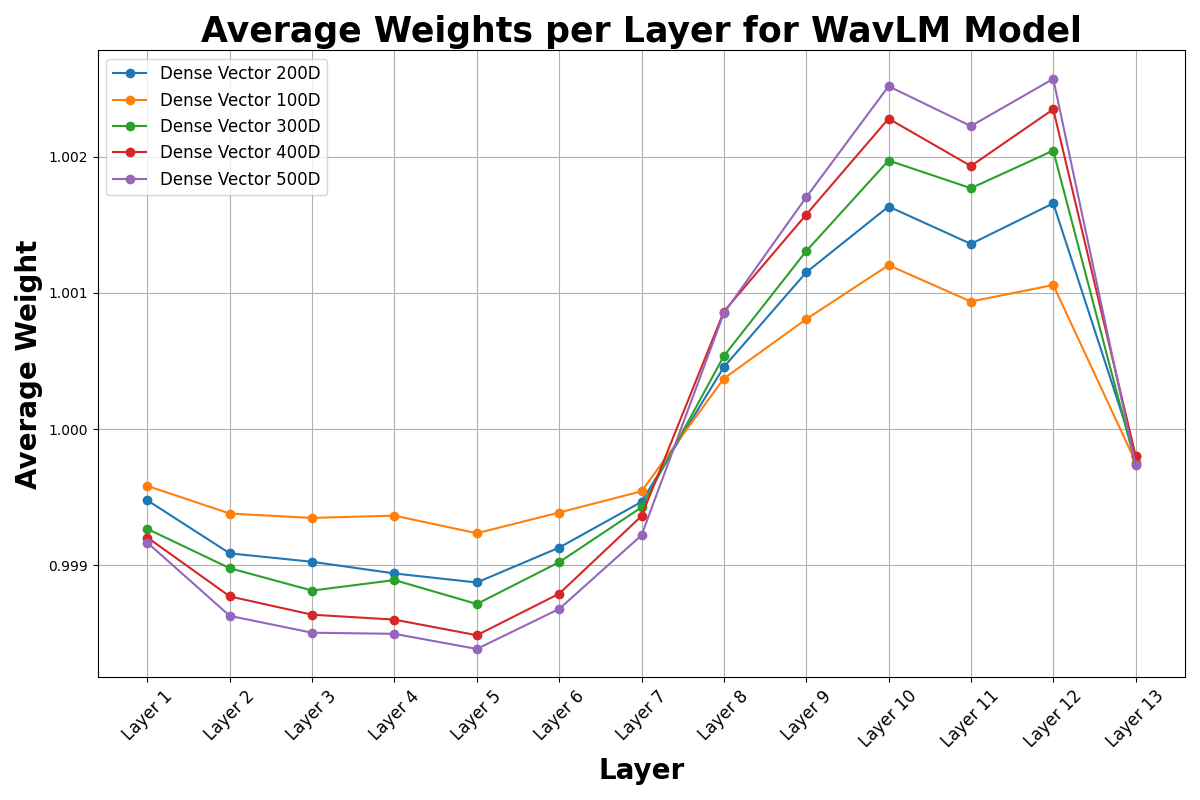}
        \caption{}
        \label{fig:image1}
    \end{subfigure}
    \hfill
    \begin{subfigure}[b]{0.24\textwidth}
        \centering
        \includegraphics[width=\linewidth]{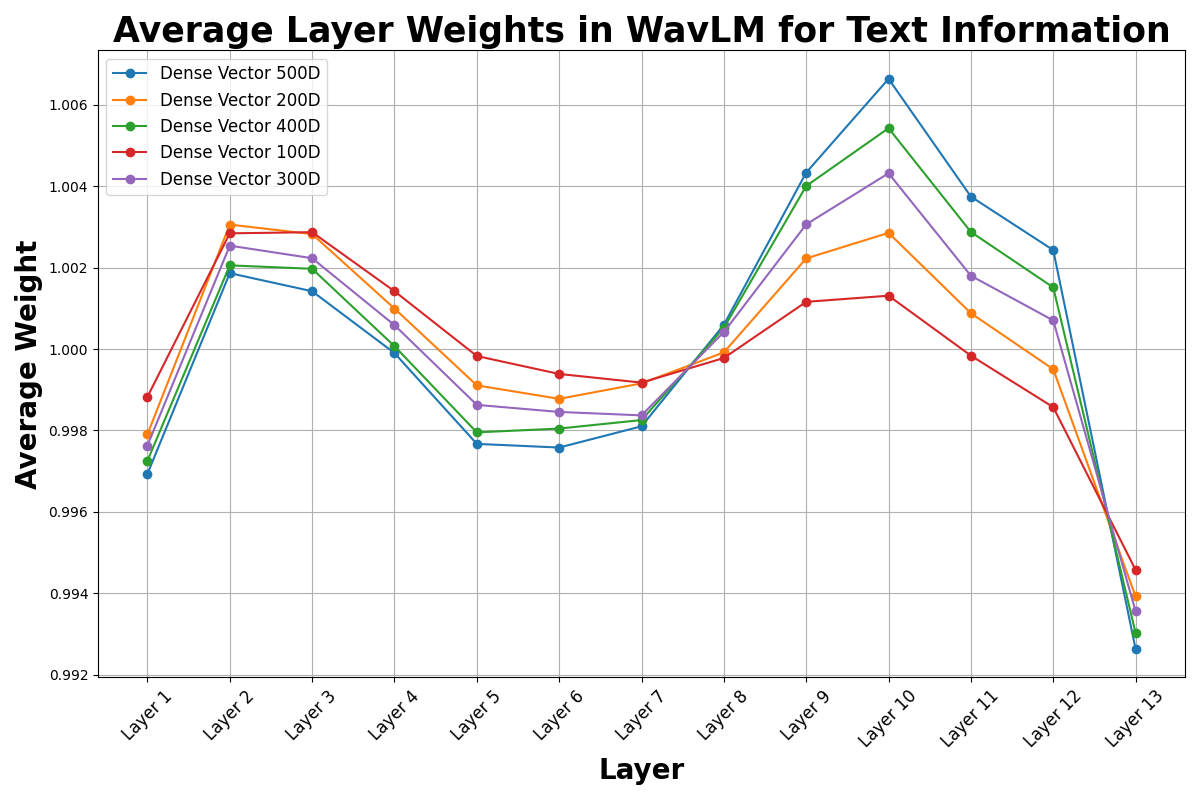}
        \caption{}
        \label{fig:image2}
    \end{subfigure}
    \hfill
    \begin{subfigure}[b]{0.24\textwidth}
        \centering
        \includegraphics[width=\linewidth]{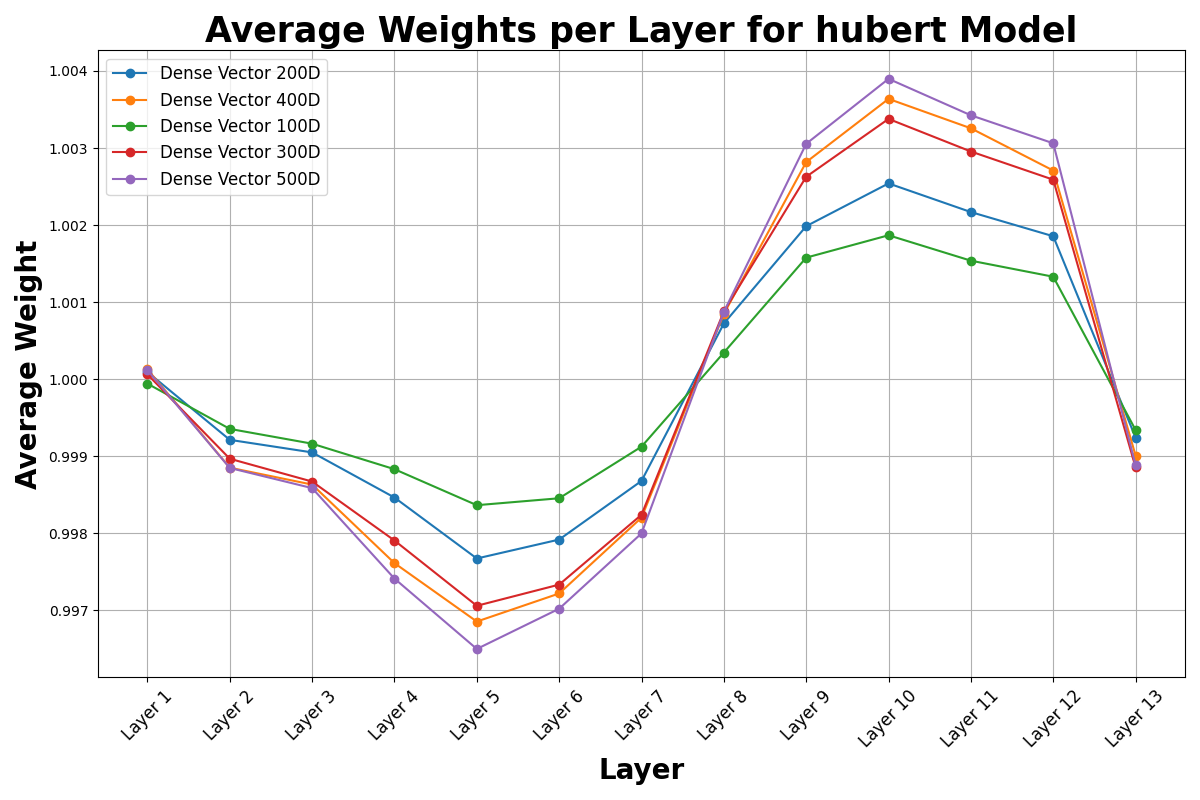}
        \caption{}
        \label{fig:image3}
    \end{subfigure}
    \hfill
    \begin{subfigure}[b]{0.24\textwidth}
        \centering
        \includegraphics[width=\linewidth]{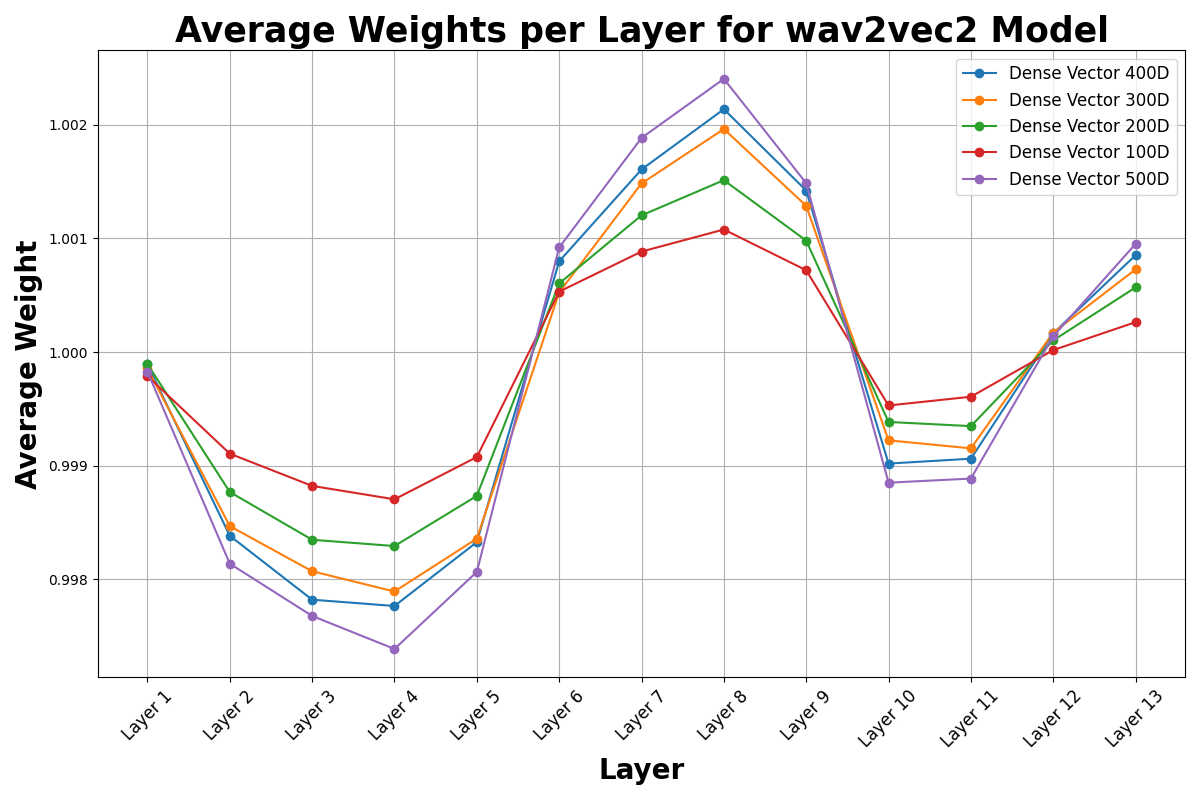}
        \caption{}
        \label{fig:image4}
    \end{subfigure}
    \caption{Average weight of different speech SSL models When learning dense vector}
    \vspace{-14pt}
    \label{fig:four_images}
\end{figure*}

Building on the insights from the information separation analysis in the previous section, we delve deeper into how layer-wise representations contribute to this separation process.~Since WavLM consistently achieves the best performance across all speech SSL models, our analysis primarily focuses on WavLM. Figure~\ref{fig:image1} reveals that layer weights for extracting speech-specific information via contrastive learning are predominantly concentrated in the deeper layers, which are known to encode high-level semantic and emotional information~\cite{chen2022wavlm} encompassing both content-related and speech-specific features. By leveraging Sentence-BERT embeddings, the contrastive learning framework disentangles these features, isolating speech-specific information while filtering out content-related aspects, reducing the overlap between the two and validating our earlier hypothesis.

Interestingly, Figure~\ref{fig:image2} demonstrates a different pattern for extracting text-specific information, with weights distributed across both earlier and deeper layers, peaking around Layer 2 and Layer 10. This pattern suggests that text-specific information leverages features across a broader range of layers, reflecting its inherently mixed nature within speech SSL models. The consistency of these trends across dense vector dimensions underscores the robustness of these observations and highlights the targeted nature of contrastive learning in shaping layer-wise representations to align with specific objectives.

The layer weight distributions for HuBERT and Wav2Vec2, as shown in Figure~\ref{fig:image3} and Figure~\ref{fig:image4}, reveal the impact of pre-training methodologies on the representation of speech-specific information. HuBERT’s distribution closely resembles WavLM’s, with weights concentrated in the deeper layers, peaking around layers 10 to 12. By contrast, Wav2Vec2 displays a more distributed pattern, with less reliance on deeper layers. This contrast highlights the role of training strategies in shaping information distribution. HuBERT and WavLM share similar masked prediction tasks on discretized pseudo-labels during pre-training, which likely drives their deeper layers to encode nuanced speech-specific features. Meanwhile, Wav2Vec2’s contrastive learning approach results in a different allocation of information across layers. These findings suggest that pre-training methodology fundamentally influences how speech and content information are organized within the model, shaping its suitability for downstream tasks like depression detection.

\subsection{How the Model Detects Depression by Dense Vector?}

Understanding how a depression detection system makes decisions is essential, as it ensures transparency and fosters trust in its outcomes. In our baseline framework, we introduced a depression detection system grounded in acoustic landmarks, with each landmark offering linguistically interpretable insights. Building on this approach, we extended the system to extract temporal information from speech using acoustic landmarks, aiming to enhance the interpretability of the dense vector depression detection system.

Our approach analyzes the most influential sentences in each file based on their impact on classification decisions. We select five sentences per file, following established practices in interpretable machine learning that suggest a small subset of high-impact examples provides better interpretability than analyzing all instances~\cite{koh2017understanding, arras2017relevant}. This focused analysis allows us to identify clear patterns in how our model makes decisions while maintaining tractable computational complexity. The five-sentence threshold balances between capturing sufficient variation in speech patterns and maintaining clear interpretability of results.

\textbf{Identifying Key Sentences.} To identify the most important sentences in a document for depression detection, we analyze how the removal of each sentence affects the SVM classifier's decision value. Given a document-level dense vector \( \mathbf{v}_{\text{doc}} \), represented as the average of sentence-level dense vectors \( \{\mathbf{v}_1, \mathbf{v}_2, \dots, \mathbf{v}_n\} \):
\begin{equation}
\mathbf{v}_{\text{doc}} = \frac{1}{n} \sum_{i=1}^n \mathbf{v}_i,
\end{equation}
the decision value for the SVM classifier is computed as:
\begin{equation}
D_{\text{original}} = f(\mathbf{v}_{\text{doc}}),
\end{equation}
where \( f(\cdot) \) is the decision function of the SVM.

For each sentence \( \mathbf{v}_i \), we construct a modified document-level vector \( \mathbf{v}_{\text{modified}} \) by excluding the \( i \)-th sentence:
\begin{equation}
\mathbf{v}_{\text{modified}} = \frac{1}{n-1} \sum_{j \neq i} \mathbf{v}_j.
\end{equation}
The modified decision value is calculated as:
\begin{equation}
D_{\text{modified}, i} = f(\mathbf{v}_{\text{modified}}).
\end{equation}

The importance of the \( i \)-th sentence is quantified as the absolute change in the decision value:
\begin{equation}
\Delta D_i = \left| D_{\text{original}} - D_{\text{modified}, i} \right|.
\end{equation}

The sentences are ranked based on their \( \Delta D_i \) values, with the top five sentences selected as the most influential for the model's prediction. For each document, this procedure also provides insight into which sentences most significantly contribute to the classification decision.

\textbf{Landmark Bigram Duration Analysis} After identifying the most influential sentences, we aim to explore their common characteristics and uncover the distinguishing features that the machine learning model leverages to differentiate between healthy individuals and those with depression. Previous research has demonstrated significant differences in temporal characteristics~\cite{cummins2015review,huang2019investigation}, such as speaking rate and pauses, between individuals with depression and healthy individuals. The acoustic landmarks in our baseline system provide a promising approach for analyzing the temporal structure of speech signals. More precisely, we derived several statistical features from the time durations between consecutive acoustic landmarks.

Each important sentence was annotated with a sequence of landmarks. For each landmark \( l_i \), its timestamp was denoted by \( t(l_i) \). The duration between two adjacent landmarks \( l_i \) and \( l_{i+1} \) was defined as:
\begin{equation}
d_{i \rightarrow i+1} = t(l_{i+1}) - t(l_i),
\end{equation}
where \( i \in \{1, \dots, n-1\} \) and \( n \) is the total number of landmarks in a given sentence.

To quantify the temporal characteristics of speech in depressed and healthy individuals, we calculated statistical features for the duration of each landmark bigram. For a given landmark bigram~\cite{huang2019investigation,huang2019natural} $b_{n \rightarrow m}$, where $n,m \in (g,p,s,f,v,b)$ its durations are aggregated across all samples, forming a set \( D_{b_{n \rightarrow m}} = \{d^1_{n \rightarrow m}, d^2_{n \rightarrow m}, \dots, d^y_{n \rightarrow m}\} \). The \( k \)-th statistical feature for this set was computed as:
\begin{equation}
d_k = \big|  D_{b_{n \rightarrow m}} \big|_k,
\end{equation}
where \( \big| \cdot \big|_k \) represents a specific statistical measure.

We calculated a series of features, including mean, median, variance, standard deviation, minimum, maximum interquartile range, skewness and kurtosis, to summarize the temporal dynamics of each bigram. These features collectively describe the differences in timing patterns between healthy and depressed individuals.

To evaluate the statistical significance of these differences, we conducted a Mann-Whitney U test for each bigram \( b_{n \rightarrow m} \) to compare the duration distributions between the two groups. The test produces a U-statistic and a p-value, indicating whether the differences are statistically significant. Bigram pairs with p-values below the threshold of 0.05 are identified as having significant differences, highlighting the landmark transitions most relevant to distinguishing depressed and healthy individuals.

\textbf{Results and Interpretation} Our analysis identified two landmark bigrams~\texttt{b--g+} and \texttt{p+-b-} as exhibiting the most significant differences in duration distributions between healthy and depressed individuals when compared with other landmark pairs. From a statistical perspective, these findings confirm that there are observable differences in speech timing characteristics between healthy individuals and those with depression. These distinctions in landmark duration patterns provide evidence that speech timing information carries meaningful variations between the two groups, aligning with the model’s ability to leverage such acoustic features for classification.

The~\texttt{b-g+} bigram captures the transition from the cessation of turbulent noise (\texttt{b-}) to the onset of vocal fold vibration (\texttt{g+}). This transition often corresponds to the release of a voiced obstruent, a feature influenced by both articulation precision and timing~\cite{nathan1998sounds}. Similarly, the \texttt{p+-b-} bigram represents the shift from the onset of periodicity (\texttt{p+})—indicative of voiced sound initiation—to the offset of turbulent noise (\texttt{b-}), which is associated with the termination of an obstruent region~\cite{nathan1998sounds}.

The prominence of these landmark pairs highlights that depression may affect the fine-grained timing and coordination of speech production mechanisms, particularly in transitions involving voicing and turbulence. Such temporal shifts could be related to changes in motor control, vocal fold tension, or articulatory effort, which have been observed in prior studies on speech characteristics of individuals with depression~\cite{williamson2013vocal,quatieri2012vocal}. These findings reinforce the importance of leveraging temporal acoustic features for uncovering depression-specific speech patterns.

\section{Large Language Model Results and Analysis}\label{LLM SECTION}

\subsection{LLM Experiment Setup}
To ensure a comprehensive comparison with previous work, we conducted experiments on four models: Llama2-7B, Llama2-7B Chat, Llama2-13B, and Llama2-13B Chat. All models were frozen to extract embeddings without fine-tuning. The experiments were performed on an NVIDIA A100 GPU with 80GB of memory. The downstream classification model used is consistent with the SVM setup described earlier.
\subsection{LLM Experiment Results}
\begin{table}[t]
    \centering
    \setlength{\tabcolsep}{3pt} 
    \renewcommand{\arraystretch}{1.1} 
    \caption{F1 Scores across different Large Language Models, embedding dimensions. Dep Info indicates if contains a hint of depression information.}
    \label{tab:llm_f1_scores}
    \scriptsize 
    \resizebox{\columnwidth}{!}{ 
    \begin{tabular}{@{}lcccccc@{}}
    \toprule
    \toprule
    \textbf{Model} & \textbf{Dim.} & \textbf{Dep. Info} & \textbf{F1-avg} & \textbf{F1-max} & \textbf{F1-std} \\
    \midrule
    \midrule
    \textbf{Llama2 7B Chat} & 300 & \xmark & 0.5259 & 0.5366 & 0.0072 \\
                            & 300 & \cmark & 0.6812 & \textbf{0.6957} & 0.0167 \\
                            & 500 & \xmark & 0.6191 & 0.6667 & 0.0550 \\
                            & 500 & \cmark & 0.6202 & 0.6364 & 0.0325 \\
    \midrule
    \textbf{Llama2 7B}      & 300 & \xmark & 0.6667 & 0.6667 & 0.0000 \\
                            & 300 & \cmark & 0.6667 & 0.6667 & 0.0000 \\
                            & 500 & \xmark & 0.6618 & 0.6667 & 0.0098 \\
                            & 500 & \cmark & 0.6667 & 0.6667 & 0.0000 \\

    \midrule
    \textbf{Llama2 13B Chat} & 300 & \xmark & 0.629 & 0.6667 & 0.0429 \\
                            & 300 & \cmark  & 0.667 & 0.6667 & 0.0000 \\
                            & 500 & \xmark & 0.6300 & 0.6400 & 0.0200 \\
                            & 500 & \cmark & 0.6600 & 0.6667 & 0.0134 \\
    \midrule
    \textbf{Llama2 13B}     & 300 & \xmark & 0.5976 & 0.6286 & 0.0449 \\
                            & 300 & \cmark & 0.6270 & 0.6667 & 0.0282 \\
                            & 500 & \xmark & 0.5786 & 0.6000 & 0.0143 \\
                            & 500 & \cmark & 0.6559 & 0.6667 & 0.0124 \\
    \midrule
    \textbf{Previous SOTA~\cite{wu2023self}} & - & - & 0.756 & \textbf{0.800} & 0.023\\
    \textbf{Chat 7B + Wavlm}& 300 & \cmark & \textbf{0.76655} & 0.7692 & 0.0053 \\
    \bottomrule
    \vspace{-5mm}
    \end{tabular}
    }
\end{table}
\begin{figure*}[t]
\centering
\begin{subfigure}[b]{0.24\textwidth}
    \centering
    \includegraphics[width=\linewidth]{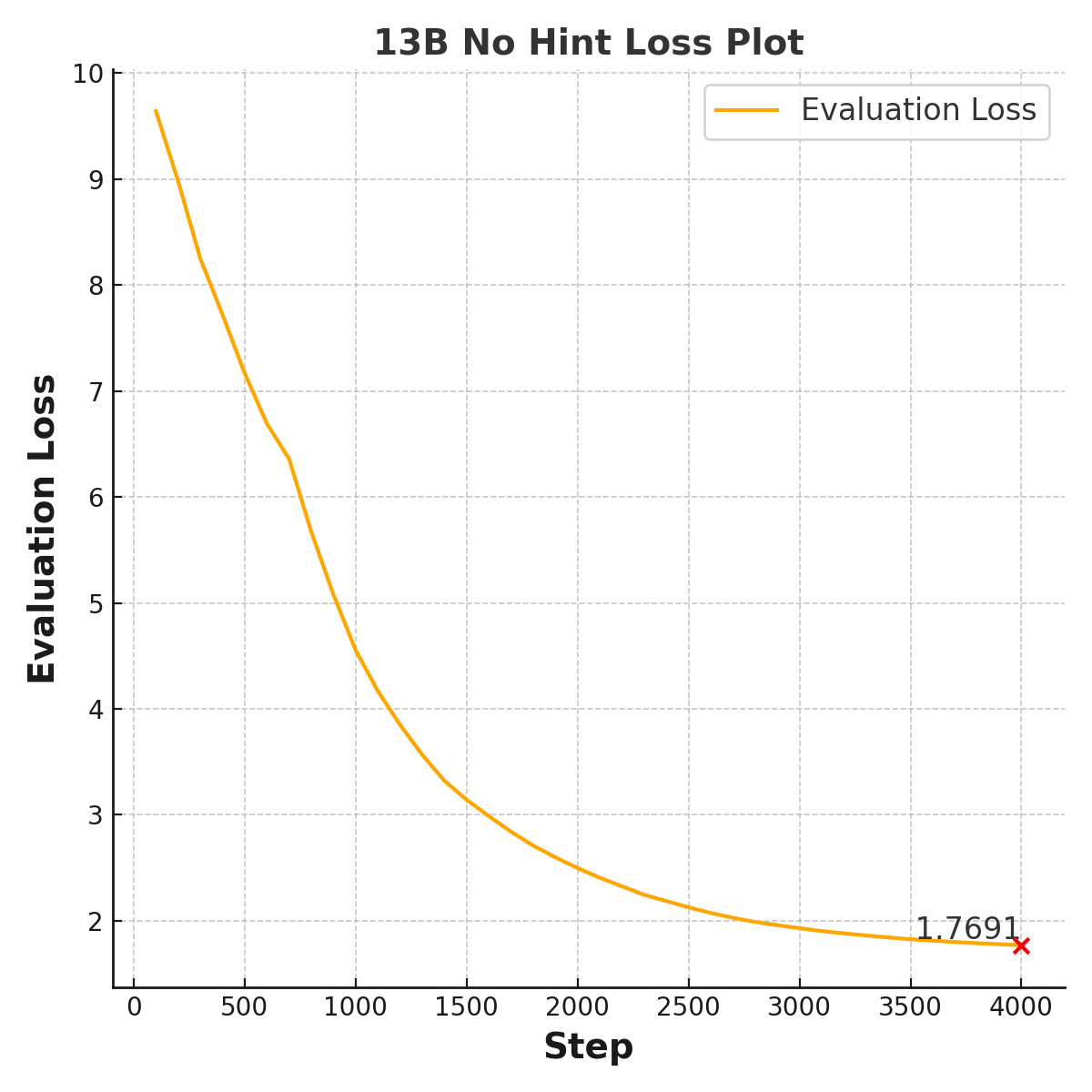}
    \caption{}
    \label{fig:13b_no_hint}
\end{subfigure}
\hfill
\begin{subfigure}[b]{0.24\textwidth}
    \centering
    \includegraphics[width=\linewidth]{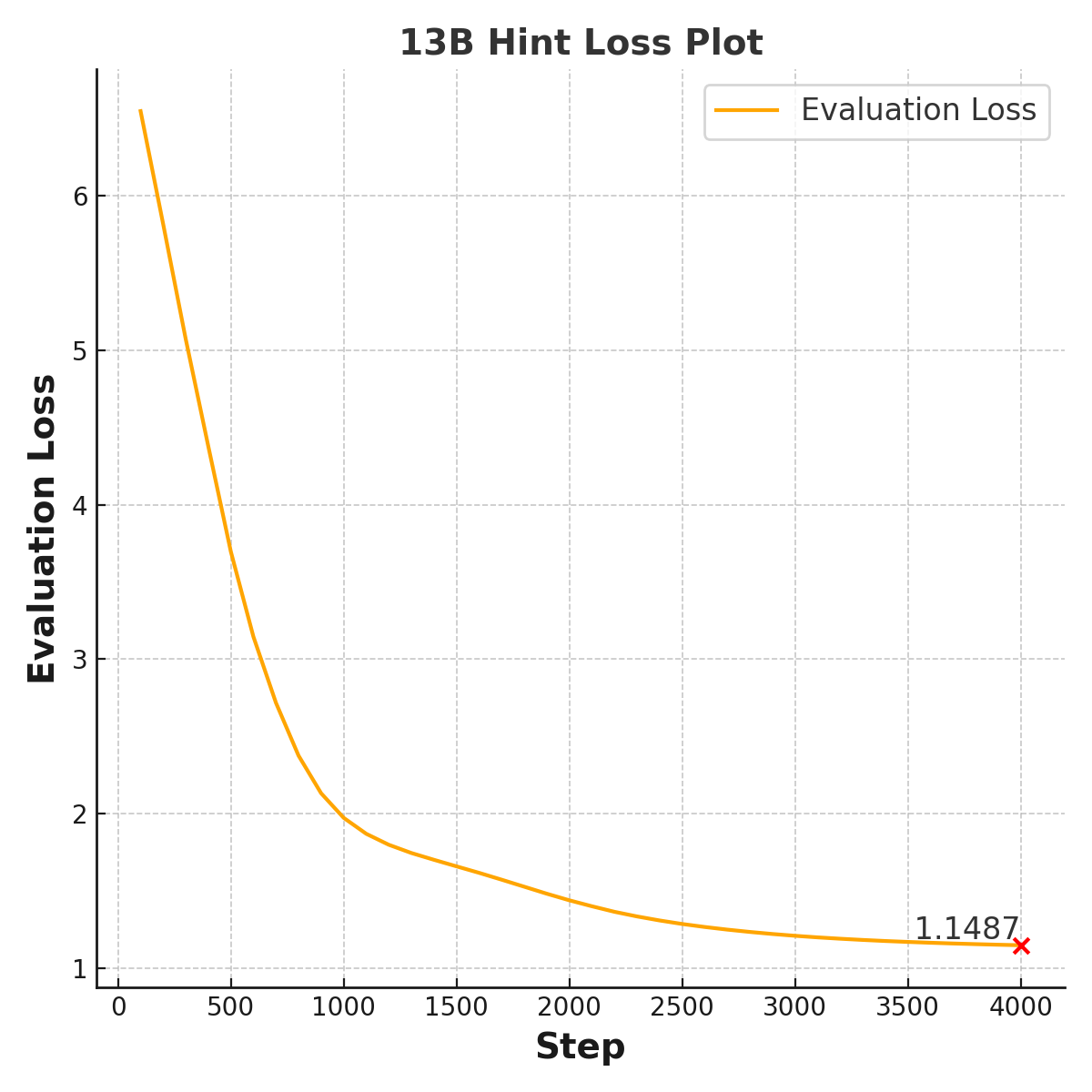}
    \caption{}
    \label{fig:13b_hint}
\end{subfigure}
\hfill
\begin{subfigure}[b]{0.24\textwidth}
    \centering
    \includegraphics[width=\linewidth]{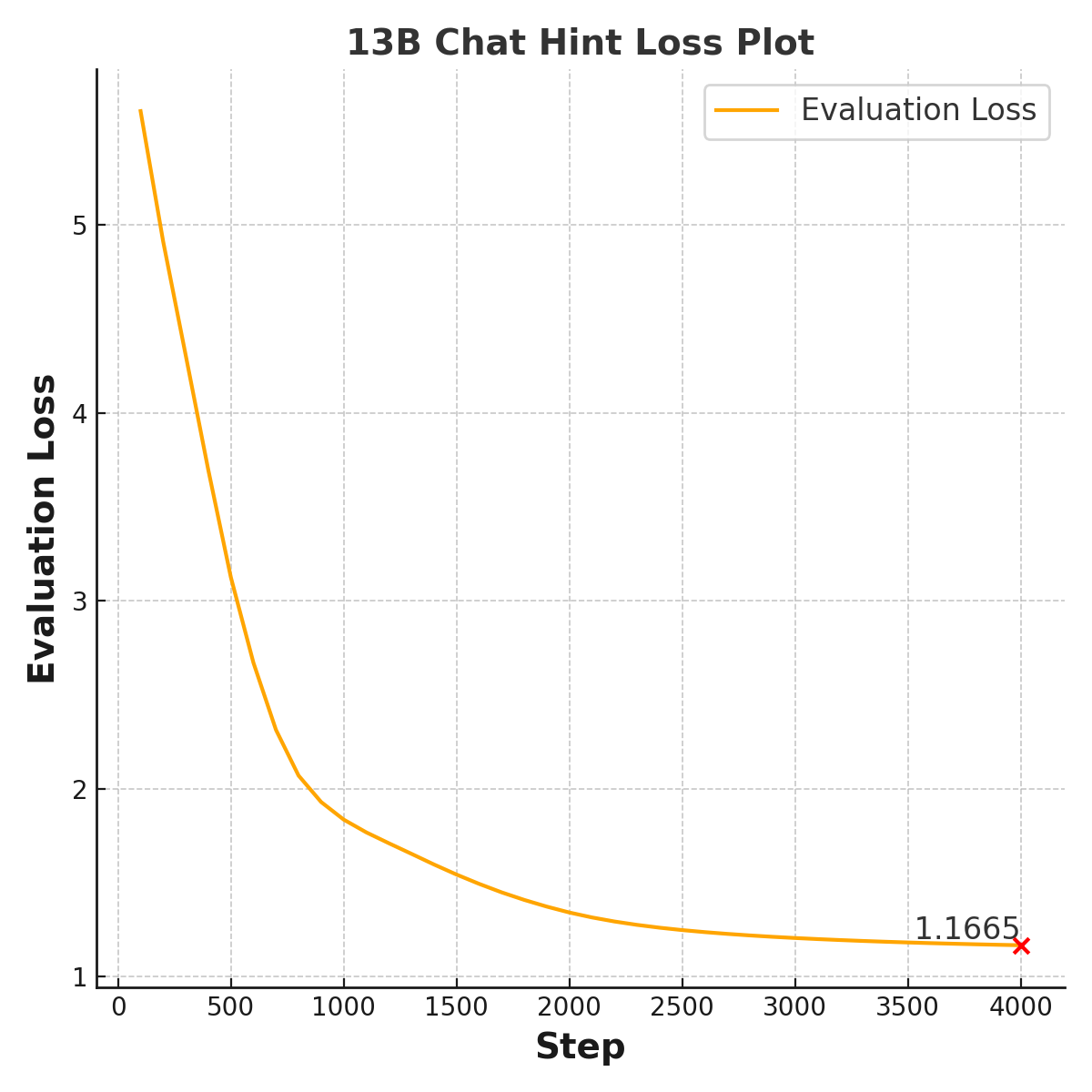}
    \caption{}
    \label{fig:13b_chat_hint}
\end{subfigure}
\hfill
\begin{subfigure}[b]{0.24\textwidth}
    \centering
    \includegraphics[width=\linewidth]{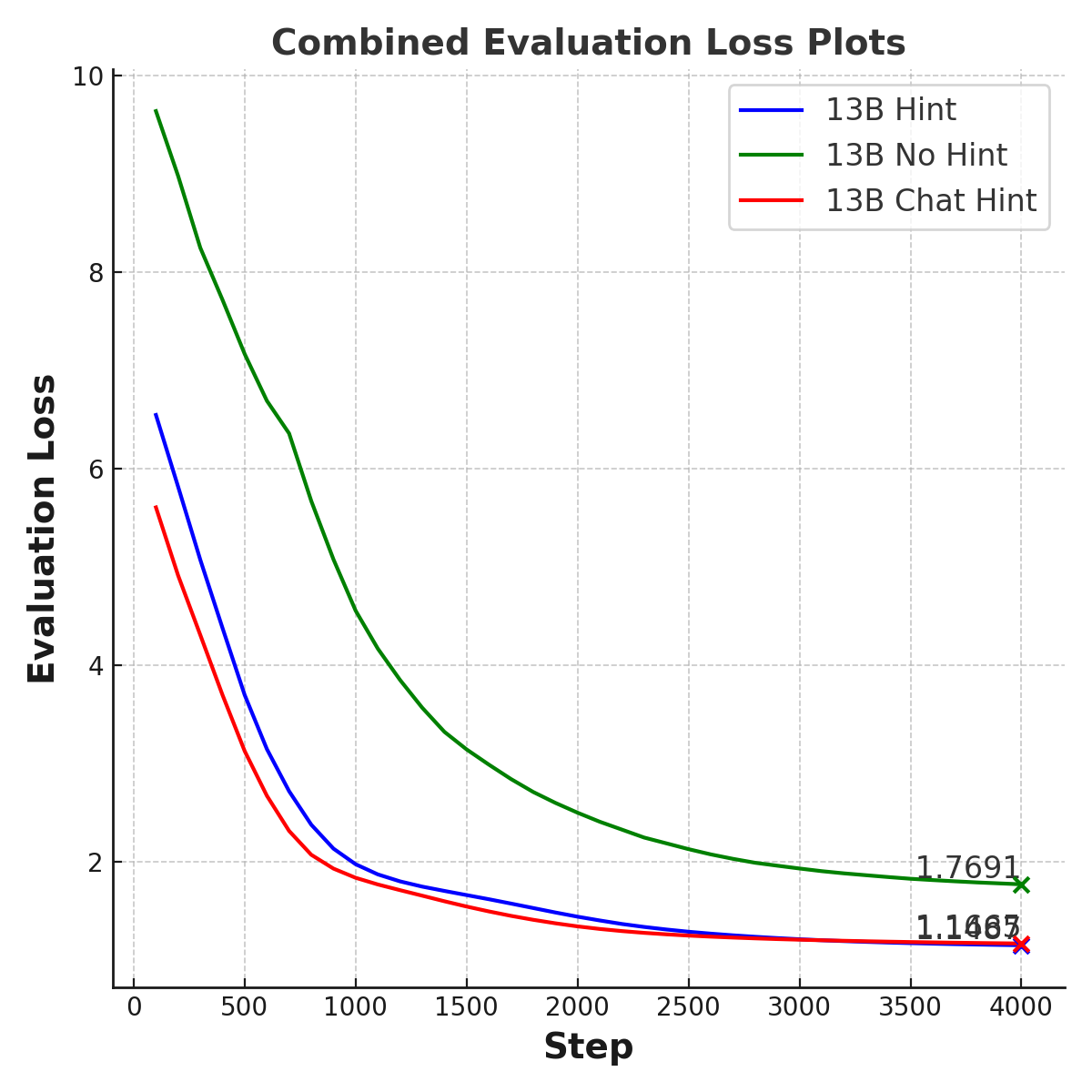}
    \caption{}
    \label{fig:combined}
\end{subfigure}
\caption{Evaluation loss for different configurations up to 4000 Steps for baseline cross-model instruction tuning.}
\label{fig:eval_loss}
\end{figure*}
Table~\ref{tab:llm_f1_scores} summarizes the F1 scores across various Llama2-based models with different embedding dimensions and the inclusion or exclusion of depression-related information (denoted as "Dep. Info").

From the results, we observe that models incorporating depression information consistently outperform their counterparts without it. For example, the Llama2 7B Chat model achieves a notable increase in F1-avg from 0.5259 (without depression information) to 0.6812 (with depression information) at the 300-dimensional setting, with a similar trend observed for other dimensions. This highlights the importance of including depression-specific information in embeddings to improve classification performance.

Among the models, Llama2 7B exhibited high consistency in their F1 scores. The Llama2 7B model achieves identical F1-avg and F1-max values of 0.6667 across all configurations, regardless of embedding dimensions or the inclusion of depression-related information, with F1-std values consistently near zero. This phenomenon indicates that the embeddings of Llama2 7B lack diversity, which might be attributed to the absence of reinforcement learning from human feedback (RLHF). In contrast, the Llama2 7B Chat model also demonstrates strong performance, but its F1-avg values exhibit greater variability across configurations, such as 0.5259 for 300 dimensions without depression information compared to 0.6812 with depression information. This variability underscores the influence of RLHF in enhancing the output diversity of the model.

When comparing models of different scales, the Llama2 13B models do not consistently outperform their smaller counterparts (Llama2 7B). For instance, at the 500-dimensional setting with depression information, the Llama2 13B model achieves an F1-avg of 0.6559, slightly lower than the 0.6667 achieved by the Llama2 7B. Additionally, the Llama2 13B Chat model achieves an F1-avg of 0.6600 in this setting, still slightly behind the Llama2 7B. These findings indicate that increased model capacity does not always translate into superior performance.

Additionally, combining the best-performing models from each modality—Llama2 7B Chat for text-based features and WavLM for speech-based dense vectors yields the highest overall performance. This integration achieves an average F1 of 0.7665 and a max F1 of 0.769 at the 300-dimensional setting, surpassing previous state-of-the-art results that relied on WavLM Layer 10 and Roberta with extensive data augmentation. Notably, our approach achieves this performance without leveraging any form of data augmentation, demonstrating the robustness and effectiveness of combining complementary features from specialized text and speech models.

\subsection{The Critical Role of Hints in Baseline and Advanced Systems}

The inclusion of hints—providing information about whether a data sample originates from a depressed or healthy individual—has demonstrated significant value across both the baseline system and the new LLM-based systems. While the mechanisms differ between the two setups, the impact of embedding task-specific information remains evident in both.

In the baseline system, hints are directly integrated into the Cross-modal Instruction Training process, and their influence is clearly illustrated by the evaluation loss trends shown in Figure~\ref{fig:eval_loss}. When no hint was provided, the loss converged to a higher value of approximately 1.76 (Figure~\ref{fig:13b_no_hint}). By contrast, the presence of hints led to a significant improvement, with the loss consistently converging to around 1.1 (Figures~\ref{fig:13b_hint} and~\ref{fig:13b_chat_hint}). Figure~\ref{fig:combined} further emphasizes this disparity, demonstrating the role of hints in driving better convergence behavior. These results highlight that embedding depression-related information enhances the model’s ability to differentiate between healthy and depressed speech patterns.

In the new system, which utilizes pre-trained and frozen LLM embeddings, hints are not used during LLM training but are incorporated into downstream processing to augment task-specific contextual information. Table~\ref{tab:llm_f1_scores} clearly reflects the benefit of hints across different LLM configurations. For example, in the Llama2 7B Chat model with 300-dimensional embeddings, the F1-avg improved from 0.5259 without hints to 0.6812 with hints. A similar pattern was observed across other models, such as Llama2 13B, where the F1-avg increased from 0.5786 to 0.6559 with the inclusion of depression-related information. These improvements demonstrate that even without modifying the LLMs themselves, incorporating hints into the downstream analysis substantially enhances performance.

The consistent findings across both the baseline and our advanced LLM-based systems underscore the critical role of providing explicit task-related information, such as hints, during depression detection. In the baseline system, hints enable the model to focus on relevant speech features, such as temporal and acoustic patterns, that distinguish healthy individuals from those with depression. Similarly, in the advanced LLM-based system, hints serve as explicit task signals, guiding the embeddings toward more relevant semantic spaces that align with depression classification. By clearly defining the task context, hints help LLMs prioritize task-relevant distinctions, demonstrating that even without fine-tuning, embedding task-specific information into pre-trained models significantly enhances their utility for specialized tasks like depression detection.

\subsection{LLM Embedding Diversity Analysis}
\begin{table}[t]
\centering
\setlength{\tabcolsep}{4pt} 
\renewcommand{\arraystretch}{1.2} 
\caption{Quantitative metrics for 7B and 7B chat embeddings}
\label{tab:embedding_metrics}
\scriptsize
\resizebox{\columnwidth}{!}{ 
\begin{tabular}{@{}lcccc@{}}
\toprule
\toprule
\textbf{LlaMA2}      & \textbf{Hint} & \textbf{Cosine Mean} & \textbf{Pairwise Dist.} & \textbf{Var./Dim.} \\ 
\midrule
\textbf{7B}         & \xmark   & 0.9500 & 18.65 & 0.0457 \\
                    & \cmark   & 0.9503 & 18.57 & 0.0463 \\
 \midrule
\textbf{7B Chat}    & \xmark   & 0.9374 & 20.01 & 0.0536 \\
                    & \cmark   & 0.9373 & 19.93 & 0.0533\\
\bottomrule
\vspace{-14pt}
\end{tabular}
}
\end{table}
To understand why frozen LLM embeddings outperform fine-tuned approaches, we analyze the embedding properties of Llama2 7B and 7B Chat models. Our analysis focuses on three key metrics that characterize embedding quality and diversity: cosine similarity, pairwise distance, and variance per dimension (Table~\ref{tab:embedding_metrics}). Cosine similarity measures the alignment between embeddings by calculating the average angle between all pairs in the high-dimensional space. Higher similarity values indicate a more homogeneous embedding representation. Pairwise distance, computed as the mean Euclidean distance across embedding pairs, quantifies the degree of clustering in the embedding space, with lower values reflecting tighter clustering. Variance per dimension captures the distribution of information across the embedding space, with higher variance suggesting a more diverse and expressive representation. These metrics collectively provide a detailed understanding of the structural differences between embeddings generated by the two models.

When comparing the Chat and non-Chat versions, the embeddings generated by the Llama2 7B Chat model exhibit lower cosine similarity values, with embeddings showing less alignment compared with the Llama2 7B model. Specifically, the cosine similarity for the Llama2 7B Chat model reflects a broader diversity in embedding representations. Pairwise distances between embeddings are also larger for the Llama2 7B Chat model, indicating that its embeddings occupy a wider feature space. Additionally, the variance per dimension for the Llama2 7B Chat model is consistently higher across both configurations, reflecting a richer and more distributed representation compared to the Llama2 7B model, which shows lower variance. This observation also explains the minimal variation in F1 scores observed for the Llama2 7B model in Table~\ref{tab:llm_f1_scores}. The tightly clustered embeddings and lower variance per dimension suggest that the Llama2 7B model captures less diverse information, resulting in consistent but less adaptive performance across different configurations. In contrast, the broader representation space of the Llama2 7B Chat model enables it to better adapt to nuanced task requirements, which is reflected in its sensitivity to the inclusion of depression-related hints.

We also observed that introducing hints led to a decrease in embedding diversity across all metrics in Table~\ref{tab:embedding_metrics}. This reduction in diversity is reflected in Table~\ref{tab:llm_f1_scores}, where the inclusion of hints consistently lowers the F1-std for the same model with the same embedding dimension. This suggests that providing task-specific information helps the model generate more consistent embeddings, resulting in reduced variability in classification performance.

\subsection{\textcolor{black}{Ablation Study on Data Augmentation Dependency}}
\begin{table}[h]
\centering
\caption{Data augmentation ablation study results compared to our proposed methods}
\begin{tabular}{lccc}
\toprule
\textbf{Method} & \textbf{F1-avg} & \textbf{F1-std} & \textbf{F1-max} \\
\midrule
WavLM (our method, 300D) & 0.755 & 0.016 & 0.769 \\
\textcolor{black}{WavLM with data aug} & \textcolor{black}{0.762} & \textcolor{black}{0.014} & \textcolor{black}{0.769} \\
\midrule
LLaMA2-13B Chat (our method, 300D) & 0.6559 & 0.0124 & 0.667 \\
\textcolor{black}{LLaMA2-13B Chat with data aug} & \textcolor{black}{0.638} & \textcolor{black}{0.0452} & \textcolor{black}{0.6667} \\
\bottomrule
\end{tabular}
\label{tab:data_aug_comparison}
\end{table}
\textcolor{black}{To provide direct empirical evidence for our hypothesis about data augmentation dependency in pre-trained models, we conducted systematic ablation experiments comparing data augmentation approaches~\cite{wu2023self} with our information separation framework. We selected WavLM (300D speech-specific features) and LLaMA2-13B Chat (500D with depression information) as representative models from SSL and LLM families, respectively, based on their optimal performance configurations identified in our experiments.}

\textcolor{black}{Table~\ref{tab:data_aug_comparison} presents the comparative analysis between our approach and traditional data augmentation methods. The results demonstrate contrasting behaviors across model architectures. For WavLM, data augmentation provides only marginal improvement (F1-avg increases from 0.755 to 0.762), indicating that proper information separation largely eliminates the dependency on extensive augmentation. More significantly, LLaMA2-13B Chat exhibits performance degradation when data augmentation is applied (F1-avg decreases from 0.660 to 0.638).}

\textcolor{black}{This counterintuitive result for LLMs validates our concern that extensive data augmentation can be detrimental to depression detection performance. The performance decline suggests that artificial expansion of training samples introduces inconsistent patterns that interfere with the subtle depression-relevant features that LLMs must capture. This finding reinforces our central argument that data augmentation dependency represents a fundamental limitation rather than an effective solution for pre-trained models in depression detection tasks.}

\textcolor{black}{\subsection{Cross-Linguistic Validation on MODMA Dataset}}

\textcolor{black}{To demonstrate the cross-linguistic generalizability of our information separation framework, we conducted additional experiments on the Chinese MODMA dataset~\cite{cai2022multi}. MODMA consists of audio recordings from Chinese speakers but lacks transcript annotations, presenting both an opportunity for cross-linguistic validation and a challenge for multimodal analysis.}

\textcolor{black}{Since our approach requires both speech and text modalities, we employed OpenAI's Whisper model~\cite{radford2023robust} for automatic speech recognition to generate transcripts from the MODMA audio recordings. We then applied our information separation framework using the WavLM model for speech-specific feature extraction. Due to the absence of proper train/validation/test splits in MODMA, we used a cross-dataset evaluation protocol: training our models on DAIC-WOZ and directly testing on the entire MODMA dataset.}

\textcolor{black}{The experimental results demonstrate the effectiveness of our approach across languages. The speech-specific information separation method achieved an F1-score of 0.678, representing an 11\% improvement over the baseline WavLM features (F1-score: 0.610). This improvement pattern is consistent with our findings on the English DAIC-WOZ dataset, suggesting that the information entanglement hypothesis and our proposed separation framework are language-agnostic and generalizable across different linguistic contexts.}

\textcolor{black}{These cross-linguistic results provide additional validation that the fundamental issue of feature entanglement in pre-trained models for depression detection extends beyond English-language data. The consistent improvements achieved through information separation across both English and Chinese datasets strengthen the theoretical foundation of our approach and demonstrate its practical applicability in diverse linguistic settings.}

\subsection{CMDC as a Boundary-Case Analysis}

\begin{table}[t]
\centering
\color{black} 
\footnotesize
\caption{F1 results on DAIC-WOZ and CMDC. Values are reported as mean (std). For CMDC, we follow prior layer choices: layer 8 for W2V2 and WavLM, and layer 10 for HuBERT.}
\label{tab:cmdc_main}
\setlength{\tabcolsep}{3pt}
\resizebox{\columnwidth}{!}{%
\begin{tabular}{lccc}
\toprule
System & DAIC-noaug & DAIC+aug & CMDC \\
\midrule
W2V2           & 0.625 & 0.627 & 0.964 (0.045) \\
HuBERT         & 0.629 & 0.667 & 0.964 (0.045) \\
WavLM          & 0.647 & 0.700 & 0.887 (0.044) \\
Acoustic ctrl. & --    & --    & 0.915 (0.075) \\
\bottomrule
\end{tabular}%
}
\end{table}


\begin{table}[t]
\centering
\color{black} 
\footnotesize
\caption{CMDC results of the proposed variants. For the proposed variants, we report the best F1 obtained with classifier tuning and hidden-dimension selection, with the selected dimension shown in parentheses.}
\label{tab:cmdc_method_limit}
\setlength{\tabcolsep}{3pt}
\resizebox{\columnwidth}{!}{%
\begin{tabular}{lccc}
\toprule
Method & W2V2 & HuBERT & WavLM \\
\midrule
Frozen SSL baseline & 0.964 (0.045) & 0.964 (0.045) & 0.887 (0.044) \\
Weighted sum        & 0.486 (300)   & 0.491 (400)   & 0.495 (400)   \\
Speech preserve     & 0.459 (300)   & 0.519 (300)   & 0.514 (300)   \\
Text preserve       & 0.451 (200)   & 0.451 (200)   & 0.451 (200)   \\
\bottomrule
\end{tabular}%
}
\end{table}

\textcolor{black}{The CMDC results should be interpreted distinctly from the MODMA results. While MODMA serves as a cross-linguistic validation set for our main hypothesis, CMDC instantiates a markedly different evaluation regime. On CMDC, the strongest frozen SSL baselines reach an F1 of 0.964—far exceeding the range observed on DAIC-WOZ and substantially higher than what typical cross-corpus variation would explain. This suggests that CMDC contains highly accessible, label-correlated global structures that dominate the prediction task.}

\textcolor{black}{Crucially, exploiting this separability does not require sophisticated representation learning. To demonstrate this, we evaluated a simple control system utilizing only low-level acoustic summary statistics. For each utterance, we extracted three basic features from the raw waveform: duration, RMS energy, and silence ratio. We then aggregated these to the subject level (using mean, standard deviation, minimum, and maximum across all utterances), appended the total utterance count, and trained a z-normalized RBF-SVM. Remarkably, this simple control achieves an F1 of 0.915 and an accuracy of 0.944 on the official folds. This performance is comparable to, and slightly exceeds, the F1 of 0.9053 reported for wav2vec 2.0 transfer learning on CMDC by~\cite{zhang2024improving}, and it leaves only a relatively narrow margin to recent, much stronger multimodal systems, such as the 0.9708 reported by~\cite{xu2025depression} and the 0.9818 reported by DepressInstruct~\cite{li2025depressinstruct}. This fundamentally shifts the interpretation of the benchmark: the near-perfect scores on CMDC likely reflect corpus-specific acoustic regularities rather than deep, depression-relevant semantic-acoustic modeling.}

\textcolor{black}{Within this context, the behavior of our disentanglement pipeline provides valuable insights into its inductive biases. While it is true that all information-separation variants yield overall F1 scores (0.451–0.519) significantly below the plain SSL baseline—a natural consequence of deep disentanglement acting as a strict filter that strips away the superficial acoustic shortcuts—a closer examination of these isolated variants reveals a compelling pattern. Specifically, retaining only the speech-specific information consistently outperforms both the semantic-only and the previously entangled representations.}

\textcolor{black}{For an analytical study investigating why pre-trained models fail, this internal comparison serves as critical empirical evidence. It demonstrates that the degradation in plain pre-trained models does not stem from an inability to capture acoustic cues. Rather, because the dominant predictive signals in CMDC are inherently acoustic, forcing the model to mix these robust speech signals with semantic features inevitably introduces confounding noise that dilutes the acoustic decision boundary. The superiority of the isolated speech representation over the mixed modalities perfectly corroborates our core thesis: it is the destructive entanglement of modalities that fundamentally hinders multi-modal depression detection.}

\textcolor{black}{Taken together, the findings from MODMA and CMDC offer a comprehensive view of our analysis. MODMA confirms the cross-linguistic robustness of our disentanglement strategy, whereas CMDC illustrates its boundary conditions. When a task can be largely solved by basic acoustic profiles, rigorous information separation naturally suppresses superficial global cues. Yet, even under such stress-test conditions, the supremacy of the isolated speech representation validates our fundamental claim: disentangling features is essential to prevent semantic noise from corrupting robust acoustic signals.}

\section{Conclusion}
In this study, we investigated why pre-trained models struggle with multi-modal depression detection and proposed a solution through information separation. Our investigation revealed that the poor performance of these models stems from the entanglement of content and speech-specific features, rather than limitations in model capacity. By systematically testing this hypothesis, we demonstrated that explicit separation of speech and content information significantly improves performance without requiring data augmentation. For SSL models, our information separation framework achieved state-of-the-art results by disentangling modality-specific features, while for LLMs, we found that frozen embeddings with appropriate prompting outperform fine-tuning approaches, suggesting that preserving pre-trained knowledge while guiding feature extraction is crucial. Our findings contribute both theoretical understanding and practical solutions to multi-modal depression detection, demonstrating that careful consideration of how information is represented and processed is essential for effective depression detection systems.

\section*{Acknowledgement}
This work was supported by the Australian Research Council Discovery Project DP230101184.

\bibliographystyle{IEEEtran}
\bibliography{reference}

\begin{thebibliography}{10}
\providecommand{\url}[1]{#1}
\csname url@rmstyle\endcsname
\providecommand{\newblock}{\relax}
\providecommand{\bibinfo}[2]{#2}
\providecommand\BIBentrySTDinterwordspacing{\spaceskip=0pt\relax}
\providecommand\BIBentryALTinterwordstretchfactor{4}
\providecommand\BIBentryALTinterwordspacing{\spaceskip=\fontdimen2\font plus
\BIBentryALTinterwordstretchfactor\fontdimen3\font minus
  \fontdimen4\font\relax}
\providecommand\BIBforeignlanguage[2]{{%
\expandafter\ifx\csname l@#1\endcsname\relax
\typeout{** WARNING: IEEEtran.bst: No hyphenation pattern has been}%
\typeout{** loaded for the language `#1'. Using the pattern for}%
\typeout{** the default language instead.}%
\else
\language=\csname l@#1\endcsname
\fi
#2}}

\bibitem{walker2018prevalence}
J.~Walker, K.~Burke, M.~Wanat, R.~Fisher, J.~Fielding, A.~Mulick, S.~Puntis,
  J.~Sharpe, M.~Degli~Esposti, E.~Harriss, \emph{et~al.}, ``The prevalence of
  depression in general hospital inpatients: a systematic review and
  meta-analysis of interview-based studies,'' \emph{Psychological medicine},
  vol.~48, no.~14, pp. 2285--2298, 2018.

\bibitem{baevski2020wav2vec}
A.~Baevski, Y.~Zhou, A.~Mohamed, and M.~Auli, ``wav2vec 2.0: A framework for
  self-supervised learning of speech representations,'' \emph{Advances in
  neural information processing systems}, vol.~33, pp. 12\,449--12\,460, 2020.

\bibitem{chen2022wavlm}
S.~Chen, C.~Wang, Z.~Chen, Y.~Wu, S.~Liu, Z.~Chen, J.~Li, N.~Kanda,
  T.~Yoshioka, X.~Xiao, \emph{et~al.}, ``Wavlm: Large-scale self-supervised
  pre-training for full stack speech processing,'' \emph{IEEE Journal of
  Selected Topics in Signal Processing}, vol.~16, no.~6, pp. 1505--1518, 2022.

\bibitem{wu2023self}
W.~Wu, C.~Zhang, and P.~C. Woodland, ``Self-supervised representations in
  speech-based depression detection,'' in \emph{ICASSP 2023-2023 IEEE
  International Conference on Acoustics, Speech and Signal Processing
  (ICASSP)}.\hskip 1em plus 0.5em minus 0.4em\relax IEEE, 2023, pp. 1--5.

\bibitem{bao2023somatisation}
Z.~Bao, K.~Qian, Z.~Zhao, M.~Sun, R.~Huang, D.~Xu, B.~Hu, Y.~Yamamoto, and
  B.~W. Schuller, ``Somatisation disorder detection via speech: Introducing a
  self-supervised learning model,'' in \emph{2023 45th Annual International
  Conference of the IEEE Engineering in Medicine \& Biology Society
  (EMBC)}.\hskip 1em plus 0.5em minus 0.4em\relax IEEE, 2023, pp. 1--4.

\bibitem{chowdhery2023palm}
A.~Chowdhery, S.~Narang, J.~Devlin, M.~Bosma, G.~Mishra, A.~Roberts, P.~Barham,
  H.~W. Chung, C.~Sutton, S.~Gehrmann, \emph{et~al.}, ``Palm: Scaling language
  modeling with pathways,'' \emph{Journal of Machine Learning Research},
  vol.~24, no. 240, pp. 1--113, 2023.

\bibitem{touvron2023llama}
H.~Touvron, L.~Martin, K.~Stone, P.~Albert, A.~Almahairi, Y.~Babaei,
  N.~Bashlykov, S.~Batra, P.~Bhargava, S.~Bhosale, \emph{et~al.}, ``Llama 2:
  Open foundation and fine-tuned chat models,'' \emph{arXiv preprint
  arXiv:2307.09288}, 2023.

\bibitem{oh2023chatgpt}
N.~Oh, G.-S. Choi, and W.~Y. Lee, ``Chatgpt goes to the operating room:
  evaluating gpt-4 performance and its potential in surgical education and
  training in the era of large language models,'' \emph{Annals of Surgical
  Treatment and Research}, vol. 104, no.~5, p. 269, 2023.

\bibitem{wang2023can}
Z.~Wang, R.~Li, B.~Dong, J.~Wang, X.~Li, N.~Liu, C.~Mao, W.~Zhang, L.~Dong,
  J.~Gao, \emph{et~al.}, ``Can llms like gpt-4 outperform traditional ai tools
  in dementia diagnosis? maybe, but not today,'' \emph{arXiv preprint
  arXiv:2306.01499}, 2023.

\bibitem{lahat2023evaluating}
A.~Lahat, E.~Shachar, B.~Avidan, Z.~Shatz, B.~S. Glicksberg, and E.~Klang,
  ``Evaluating the use of large language model in identifying top research
  questions in gastroenterology,'' \emph{Scientific reports}, vol.~13, no.~1,
  p. 4164, 2023.

\bibitem{liu1996landmark}
S.~A. Liu, ``Landmark detection for distinctive feature-based speech
  recognition,'' \emph{The Journal of the Acoustical Society of America}, vol.
  100, no.~5, pp. 3417--3430, 1996.

\bibitem{zhang2024auto}
X.~Zhang, D.~Liu, T.~Xiao, C.~Xiao, T.~Szalay, M.~Shahin, B.~Ahmed, and
  J.~Epps, ``Auto-landmark: Acoustic landmark dataset and open-source toolkit
  for landmark extraction,'' \emph{arXiv preprint arXiv:2409.07969}, 2024.

\bibitem{stevens2002toward}
K.~N. Stevens, ``Toward a model for lexical access based on acoustic landmarks
  and distinctive features,'' \emph{The Journal of the Acoustical Society of
  America}, vol. 111, no.~4, pp. 1872--1891, 2002.

\bibitem{zhang2024llms}
X.~Zhang, H.~Liu, K.~Xu, Q.~Zhang, D.~Liu, B.~Ahmed, and J.~Epps, ``When {LLM}s
  meets acoustic landmarks: An efficient approach to integrate speech into
  large language models for depression detection,'' in \emph{Proceedings of the
  2024 Conference on Empirical Methods in Natural Language Processing},
  Y.~Al-Onaizan, M.~Bansal, and Y.-N. Chen, Eds.\hskip 1em plus 0.5em minus
  0.4em\relax Miami, Florida, USA: Association for Computational Linguistics,
  Nov. 2024, pp. 146--158.

\bibitem{garvin1953preliminaries}
P.~L. Garvin, ``Preliminaries to speech analysis: The distinctive features and
  their correlates,'' 1953.

\bibitem{he2019ctc}
D.~He, X.~Yang, B.~P. Lim, Y.~Liang, M.~Hasegawa-Johnson, and D.~Chen, ``When
  ctc training meets acoustic landmarks,'' in \emph{ICASSP 2019-2019 IEEE
  International Conference on Acoustics, Speech and Signal Processing
  (ICASSP)}.\hskip 1em plus 0.5em minus 0.4em\relax IEEE, 2019, pp. 5996--6000.

\bibitem{huang2018depression}
Z.~Huang, J.~Epps, D.~Joachim, and M.~Chen, ``Depression detection from short
  utterances via diverse smartphones in natural environmental conditions.'' in
  \emph{INTERSPEECH}, 2018, pp. 3393--3397.

\bibitem{huang2019investigation}
Z.~Huang, J.~Epps, and D.~Joachim, ``Investigation of speech landmark patterns
  for depression detection,'' \emph{IEEE transactions on affective computing},
  vol.~13, no.~2, pp. 666--679, 2019.

\bibitem{boyce2012speechmark}
S.~Boyce, H.~Fell, and J.~MacAuslan, ``Speechmark: Landmark detection tool for
  speech analysis,'' in \emph{Thirteenth Annual Conference of the International
  Speech Communication Association}, 2012.

\bibitem{noble2006support}
W.~S. Noble, ``What is a support vector machine?'' \emph{Nature biotechnology},
  vol.~24, no.~12, pp. 1565--1567, 2006.

\bibitem{cummins2011investigation}
N.~Cummins, J.~Epps, M.~Breakspear, and R.~Goecke, ``An investigation of
  depressed speech detection: Features and normalization,'' in \emph{Twelfth
  Annual Conference of the International Speech Communication Association},
  2011.

\bibitem{gulati2020conformer}
A.~Gulati, J.~Qin, C.-C. Chiu, N.~Parmar, Y.~Zhang, J.~Yu, W.~Han, S.~Wang,
  Z.~Zhang, Y.~Wu, \emph{et~al.}, ``Conformer: Convolution-augmented
  transformer for speech recognition,'' \emph{arXiv preprint arXiv:2005.08100},
  2020.

\bibitem{zhang2024mamba}
X.~Zhang, Q.~Zhang, H.~Liu, T.~Xiao, X.~Qian, B.~Ahmed, E.~Ambikairajah, H.~Li,
  and J.~Epps, ``Mamba in speech: Towards an alternative to self-attention,''
  \emph{arXiv preprint arXiv:2405.12609}, 2024.

\bibitem{zhao2020hierarchical}
Z.~Zhao, Z.~Bao, Z.~Zhang, N.~Cummins, H.~Wang, and B.~Schuller, ``Hierarchical
  attention transfer networks for depression assessment from speech,'' in
  \emph{ICASSP 2020-2020 IEEE international conference on acoustics, speech and
  signal processing (ICASSP)}.\hskip 1em plus 0.5em minus 0.4em\relax IEEE,
  2020, pp. 7159--7163.

\bibitem{shen2022automatic}
Y.~Shen, H.~Yang, and L.~Lin, ``Automatic depression detection: An emotional
  audio-textual corpus and a gru/bilstm-based model,'' in \emph{ICASSP
  2022-2022 IEEE International Conference on Acoustics, Speech and Signal
  Processing (ICASSP)}.\hskip 1em plus 0.5em minus 0.4em\relax IEEE, 2022, pp.
  6247--6251.

\bibitem{hsu2021hubert}
W.-N. Hsu, B.~Bolte, Y.-H.~H. Tsai, K.~Lakhotia, R.~Salakhutdinov, and
  A.~Mohamed, ``Hubert: Self-supervised speech representation learning by
  masked prediction of hidden units,'' \emph{IEEE/ACM Transactions on Audio,
  Speech, and Language Processing}, vol.~29, pp. 3451--3460, 2021.

\bibitem{efficient_lid}
H.~Liu, L.~P.~G. Perera, A.~W.~H. Khong, E.~S. Chng, S.~J. Styles, and
  S.~Khudanpur, ``Efficient self-supervised learning representations for spoken
  language identification,'' \emph{IEEE J. Sel. Topics Signal Process.},
  vol.~16, no.~6, pp. 1296--1307, 2022.

\bibitem{liu2019roberta}
Y.~Liu, M.~Ott, N.~Goyal, J.~Du, M.~Joshi, D.~Chen, O.~Levy, M.~Lewis,
  L.~Zettlemoyer, and V.~Stoyanov, ``Roberta: A robustly optimized bert
  pretraining approach,'' \emph{arXiv preprint arXiv:1907.11692}, 2019.

\bibitem{hu2022lora}
E.~J. Hu, yelong shen, P.~Wallis, Z.~Allen-Zhu, Y.~Li, S.~Wang, L.~Wang, and
  W.~Chen, ``Lo{RA}: {L}ow-rank adaptation of large language models,'' in
  \emph{Proc. Int. Conf. Learn. Representations}, 2022.

\bibitem{li2021prefix}
X.~L. Li and P.~Liang, ``Prefix-tuning: Optimizing continuous prompts for
  generation,'' in \emph{Proceedings of the 59th Annual Meeting of the
  Association for Computational Linguistics and the 11th International Joint
  Conference on Natural Language Processing (Volume 1: Long Papers)}, 2021, pp.
  4582--4597.

\bibitem{LIU2023}
\BIBentryALTinterwordspacing
X.~Liu, Y.~Zheng, Z.~Du, M.~Ding, Y.~Qian, Z.~Yang, and J.~Tang, ``Gpt
  understands, too,'' \emph{AI Open}, 2023. [Online]. Available:
  \url{https://www.sciencedirect.com/science/article/pii/S2666651023000141}
\BIBentrySTDinterwordspacing

\bibitem{zaken2022bitfit}
E.~B. Zaken, Y.~Goldberg, and S.~Ravfogel, ``Bitfit: Simple parameter-efficient
  fine-tuning for transformer-based masked language-models,'' in
  \emph{Proceedings of the 60th Annual Meeting of the Association for
  Computational Linguistics (Volume 2: Short Papers)}, 2022, pp. 1--9.

\bibitem{montejo2024survey}
A.~Montejo-Raez, M.~D. Molina-Gonzalez, S.~M. Jimenez-Zafra, M.~A.
  Garcia-Cumbreras, and L.~J. Garcia-Lopez, ``A survey on detecting mental
  disorders with natural language processing: Literature review, trends and
  challenges,'' \emph{Computer Science Review}, vol.~53, p. 100654, 2024.

\bibitem{carlsson2019distress}
G.~K. Khazanov, C.~Xu, B.~D. Dunn, Z.~D. Cohen, R.~J. DeRubeis, and S.~D.
  Hollon, ``Distress and anhedonia as predictors of depression treatment
  outcome: A secondary analysis of a randomized clinical trial,''
  \emph{Behaviour Research and Therapy}, vol. 125, p. 103507, 2020.

\bibitem{nasir2019multimodal}
M.~Nasir, A.~Jati, P.~G. Shivakumar, S.~Nallan~Chakravarthula, and P.~Georgiou,
  ``Multimodal and multiresolution depression detection from speech and facial
  landmark features,'' in \emph{Proceedings of the 6th international workshop
  on audio/visual emotion challenge}, 2016, pp. 43--50.

\bibitem{fan2019multi}
W.~Fan, Z.~He, X.~Xing, B.~Cai, and W.~Lu, ``Multi-modality depression
  detection via multi-scale temporal dilated cnns,'' in \emph{Proceedings of
  the 9th International on Audio/Visual Emotion Challenge and Workshop}, 2019,
  pp. 73--80.

\bibitem{mamidisetti2022multimodal}
S.~Mamidisetti and M.~Reddy, ``Multimodal depression detection using audio,
  visual and textual cues: A survey,'' \emph{NeuroQuantology}, vol.~20, no.~4,
  pp. 325--338, 2022.

\bibitem{hashem2023speech}
A.~Hashem, M.~Arif, and M.~Alghamdi, ``Speech emotion recognition approaches: A
  systematic review,'' \emph{Speech Communication}, vol. 154, p. 102974, 2023.

\bibitem{fairbairn2013detecting}
Y.~Yang, C.~Fairbairn, and J.~F. Cohn, ``Detecting depression severity from
  vocal prosody,'' \emph{IEEE transactions on affective computing}, vol.~4,
  no.~2, pp. 142--150, 2012.

\bibitem{seifpanahi2023association}
M.-S. Seifpanahi, T.~Ghaemi, A.~Ghaleiha, D.~Sobhani-Rad, and M.-K. Zarabian,
  ``The association between depression severity, prosody, and voice acoustic
  features in women with depression,'' \emph{The Scientific World Journal},
  vol. 2023, no.~1, p. 9928446, 2023.

\bibitem{low2020acoustic}
J.~Wang, L.~Zhang, T.~Liu, W.~Pan, B.~Hu, and T.~Zhu, ``Acoustic differences
  between healthy and depressed people: a cross-situation study,'' \emph{BMC
  psychiatry}, vol.~19, no.~1, p. 300, 2019.

\bibitem{yang2017bio}
S.~A. Almaghrabi, S.~R. Clark, and M.~Baumert, ``Bio-acoustic features of
  depression: A review,'' \emph{Biomedical Signal Processing and Control},
  vol.~85, p. 105020, 2023.

\bibitem{etienne-etal-2024-emotion}
\BIBentryALTinterwordspacing
A.~{\'E}tienne, D.~Battistelli, and G.~Lecorv{\'e}, ``Emotion identification
  for {F}rench in written texts: Considering modes of emotion expression as a
  step towards text complexity analysis,'' in \emph{Proceedings of the 14th
  Workshop on Computational Approaches to Subjectivity, Sentiment, {\&} Social
  Media Analysis}, O.~De~Clercq, V.~Barriere, J.~Barnes, R.~Klinger, J.~Sedoc,
  and S.~Tafreshi, Eds.\hskip 1em plus 0.5em minus 0.4em\relax Bangkok,
  Thailand: Association for Computational Linguistics, Aug. 2024, pp. 168--185.
  [Online]. Available: \url{https://aclanthology.org/2024.wassa-1.14/}
\BIBentrySTDinterwordspacing

\bibitem{kim2019analysis}
E.~Kim and R.~Klinger, ``An analysis of emotion communication channels in
  fan-fiction: Towards emotional storytelling,'' in \emph{Proceedings of the
  Second Workshop on Storytelling}, 2019, pp. 56--64.

\bibitem{williamson2016detecting}
J.~R. Williamson, E.~Godoy, M.~Cha, A.~Schwarzentruber, P.~Khorrami, Y.~Gwon,
  H.-T. Kung, C.~Dagli, and T.~F. Quatieri, ``Detecting depression using vocal,
  facial and semantic communication cues,'' in \emph{Proceedings of the 6th
  international workshop on audio/visual emotion challenge}, 2016, pp. 11--18.

\bibitem{wu2022novel}
J.~Wu, T.~Dang, V.~Sethu, and E.~Ambikairajah, ``A novel markovian framework
  for integrating absolute and relative ordinal emotion information,''
  \emph{IEEE Transactions on Affective Computing}, vol.~14, no.~3, pp.
  2089--2101, 2022.

\bibitem{zhang-etal-2025-speecht}
\BIBentryALTinterwordspacing
X.~Zhang, H.~Liu, Q.~Zhang, B.~Ahmed, and J.~Epps, ``{S}peech{T}-{RAG}:
  Reliable depression detection in {LLM}s with retrieval-augmented generation
  using speech timing information,'' in \emph{Findings of the Association for
  Computational Linguistics: ACL 2025}, W.~Che, J.~Nabende, E.~Shutova, and
  M.~T. Pilehvar, Eds.\hskip 1em plus 0.5em minus 0.4em\relax Vienna, Austria:
  Association for Computational Linguistics, July 2025, pp. 10\,019--10\,030.
  [Online]. Available: \url{https://aclanthology.org/2025.findings-acl.521/}
\BIBentrySTDinterwordspacing

\bibitem{zheng2023two}
W.~Zheng, L.~Yan, and F.-Y. Wang, ``Two birds with one stone:
  Knowledge-embedded temporal convolutional transformer for depression
  detection and emotion recognition,'' \emph{IEEE Transactions on Affective
  Computing}, 2023.

\bibitem{10872825}
M.~Zhao, H.~Gao, L.~Zhao, Z.~Wang, F.~Wang, W.~Zheng, J.~Li, and C.~Liu,
  ``Decoupled multi-perspective fusion for speech depression detection,''
  \emph{IEEE Transactions on Affective Computing}, pp. 1--15, 2025.

\bibitem{li2025depressinstruct}
D.~Li, L.~Ding, Z.~Wang, and K.~Zhao, ``Depressinstruct: Instruction tuning of
  large speech-language models for depression detection,'' \emph{Information
  Fusion}, p. 104077, 2025.

\bibitem{cheng2025multisource}
L.~Zhou, Z.~Liu, Y.~Li, Y.~Duan, H.~Yu, and B.~Hu, ``Multi fine-grained fusion
  network for depression detection,'' \emph{ACM Transactions on Multimedia
  Computing, Communications and Applications}, vol.~20, no.~8, pp. 1--23, 2024.

\bibitem{niu2024depressionmlp}
M.~Niu, Y.~Li, J.~Tao, X.~Zhou, and B.~W. Schuller, ``Depressionmlp: A
  multi-layer perceptron architecture for automatic depression level prediction
  via facial keypoints and action units,'' \emph{IEEE Transactions on Circuits
  and Systems for Video Technology}, vol.~34, no.~9, pp. 8924--8938, 2024.

\bibitem{cummins2015review}
N.~Cummins, S.~Scherer, J.~Krajewski, S.~Schnieder, J.~Epps, and T.~F.
  Quatieri, ``A review of depression and suicide risk assessment using speech
  analysis,'' \emph{Speech communication}, vol.~71, pp. 10--49, 2015.

\bibitem{devault2014simsensei}
D.~DeVault, R.~Artstein, G.~Benn, T.~Dey, E.~Fast, A.~Gainer, K.~Georgila,
  J.~Gratch, A.~Hartholt, M.~Lhommet, \emph{et~al.}, ``Simsensei kiosk: A
  virtual human interviewer for healthcare decision support,'' in
  \emph{Proceedings of the 2014 international conference on Autonomous agents
  and multi-agent systems}, 2014, pp. 1061--1068.

\bibitem{gong2017topic}
Y.~Gong and C.~Poellabauer, ``Topic modeling based multi-modal depression
  detection,'' in \emph{Proceedings of the 7th annual workshop on Audio/Visual
  emotion challenge}, 2017, pp. 69--76.

\bibitem{wu2022climate}
W.~Wu, M.~Wu, and K.~Yu, ``Climate and weather: Inspecting depression detection
  via emotion recognition,'' in \emph{ICASSP 2022-2022 IEEE International
  Conference on Acoustics, Speech and Signal Processing (ICASSP)}.\hskip 1em
  plus 0.5em minus 0.4em\relax IEEE, 2022, pp. 6262--6266.

\bibitem{arcan2024assessment}
M.~Arcan, D.-P. Niland, and F.~Delahunty, ``An assessment on comprehending
  mental health through large language models,'' \emph{arXiv preprint
  arXiv:2401.04592}, 2024.

\bibitem{bergstra_algorithms_2011}
J.~Bergstra, R.~Bardenet, Y.~Bengio, and B.~K{\'e}gl, ``Algorithms for
  {{Hyper-Parameter Optimization}},'' in \emph{Advances in {{Neural Information
  Processing Systems}}}, vol.~24.\hskip 1em plus 0.5em minus 0.4em\relax
  {Curran Associates, Inc.}, 2011.

\bibitem{zhang2023llama}
R.~Zhang, J.~Han, A.~Zhou, X.~Hu, S.~Yan, P.~Lu, H.~Li, P.~Gao, and Y.~Qiao,
  ``Llama-adapter: Efficient fine-tuning of language models with zero-init
  attention,'' \emph{arXiv preprint arXiv:2303.16199}, 2023.

\bibitem{li2023prompting}
Y.~Li, Y.~Wu, J.~Li, and S.~Liu, ``Prompting large language models for
  zero-shot domain adaptation in speech recognition,'' \emph{arXiv preprint
  arXiv:2306.16007}, 2023.

\bibitem{kothaunderstanding}
S.~Kotha, J.~M. Springer, and A.~Raghunathan, ``Understanding catastrophic
  forgetting in language models via implicit inference,'' in \emph{The Twelfth
  International Conference on Learning Representations}, 2024.

\bibitem{huang2024mitigating}
J.~Huang, L.~Cui, A.~Wang, C.~Yang, X.~Liao, L.~Song, J.~Yao, and J.~Su,
  ``Mitigating catastrophic forgetting in large language models with
  self-synthesized rehearsal,'' in \emph{Proceedings of the 62nd Annual Meeting
  of the Association for Computational Linguistics (Volume 1: Long Papers)},
  2024, pp. 1416--1428.

\bibitem{gao2021simcse}
T.~Gao, X.~Yao, and D.~Chen, ``Simcse: Simple contrastive learning of sentence
  embeddings,'' in \emph{EMNLP 2021-2021 Conference on Empirical Methods in
  Natural Language Processing, Proceedings}, 2021.

\bibitem{yang2021superb}
S.-w. Yang, P.-H. Chi, Y.-S. Chuang, C.-I.~J. Lai, K.~Lakhotia, Y.~Y. Lin,
  A.~T. Liu, J.~Shi, X.~Chang, G.-T. Lin, \emph{et~al.}, ``Superb: Speech
  processing universal performance benchmark,'' \emph{arXiv preprint
  arXiv:2105.01051}, 2021.

\bibitem{mcinnes2018umap}
L.~McInnes, J.~Healy, N.~Saul, and L.~Gro{\ss}berger, ``Umap: Uniform manifold
  approximation and projection,'' \emph{Journal of Open Source Software},
  vol.~3, no.~29, p. 861, 2018.

\bibitem{belghazi2018mine}
M.~I. Belghazi, A.~Baratin, S.~Rajeshwar, S.~Ozair, Y.~Bengio, A.~Courville,
  and D.~Hjelm, ``Mutual information neural estimation,'' in
  \emph{International conference on machine learning}.\hskip 1em plus 0.5em
  minus 0.4em\relax PMLR, 2018, pp. 531--540.

\bibitem{geiger2021information}
B.~C. Geiger, ``On information plane analyses of neural network classifiers—a
  review,'' \emph{IEEE Transactions on Neural Networks and Learning Systems},
  vol.~33, no.~12, pp. 7039--7051, 2021.

\bibitem{koh2017understanding}
P.~W. Koh and P.~Liang, ``Understanding black-box predictions via influence
  functions,'' in \emph{International conference on machine learning}.\hskip
  1em plus 0.5em minus 0.4em\relax PMLR, 2017, pp. 1885--1894.

\bibitem{arras2017relevant}
L.~Arras, F.~Horn, G.~Montavon, K.-R. M{\"u}ller, and W.~Samek, ``" what is
  relevant in a text document?": An interpretable machine learning approach,''
  \emph{PloS one}, vol.~12, no.~8, p. e0181142, 2017.

\bibitem{huang2019natural}
Z.~Huang, J.~Epps, D.~Joachim, and V.~Sethu, ``Natural language processing
  methods for acoustic and landmark event-based features in speech-based
  depression detection,'' \emph{IEEE Journal of Selected Topics in Signal
  Processing}, vol.~14, no.~2, pp. 435--448, 2019.

\bibitem{nathan1998sounds}
G.~S. Nathan, ``The sounds of the world's languages,'' 1998.

\bibitem{williamson2013vocal}
J.~R. Williamson, T.~F. Quatieri, B.~S. Helfer, R.~Horwitz, B.~Yu, and D.~D.
  Mehta, ``Vocal biomarkers of depression based on motor incoordination,'' in
  \emph{Proceedings of the 3rd ACM international workshop on Audio/visual
  emotion challenge}, 2013, pp. 41--48.

\bibitem{quatieri2012vocal}
T.~F. Quatieri and N.~Malyska, ``Vocal-source biomarkers for depression: A link
  to psychomotor activity.'' in \emph{Interspeech}, vol.~2, 2012, pp.
  1059--1062.

\bibitem{cai2022multi}
H.~Cai, Z.~Yuan, Y.~Gao, S.~Sun, N.~Li, F.~Tian, H.~Xiao, J.~Li, Z.~Yang,
  X.~Li, \emph{et~al.}, ``A multi-modal open dataset for mental-disorder
  analysis,'' \emph{Scientific Data}, vol.~9, no.~1, p. 178, 2022.

\bibitem{radford2023robust}
A.~Radford, J.~W. Kim, T.~Xu, G.~Brockman, C.~McLeavey, and I.~Sutskever,
  ``Robust speech recognition via large-scale weak supervision,'' in
  \emph{International conference on machine learning}.\hskip 1em plus 0.5em
  minus 0.4em\relax PMLR, 2023, pp. 28\,492--28\,518.

\bibitem{zhang2024improving}
X.~Zhang, X.~Zhang, W.~Chen, C.~Li, and C.~Yu, ``Improving speech depression
  detection using transfer learning with wav2vec 2.0 in low-resource
  environments,'' \emph{Scientific Reports}, vol.~14, no.~1, p. 9543, 2024.

\bibitem{xu2025depression}
Z.~Xu, Y.~Gao, F.~Wang, L.~Zhang, L.~Zhang, J.~Wang, and J.~Shu, ``Depression
  detection methods based on multimodal fusion of voice and text,''
  \emph{Scientific Reports}, vol.~15, no.~1, p. 21907, 2025.

\end{thebibliography}

\begin{IEEEbiography}[{\includegraphics[width=1.1in,height=1.2in,clip,keepaspectratio]{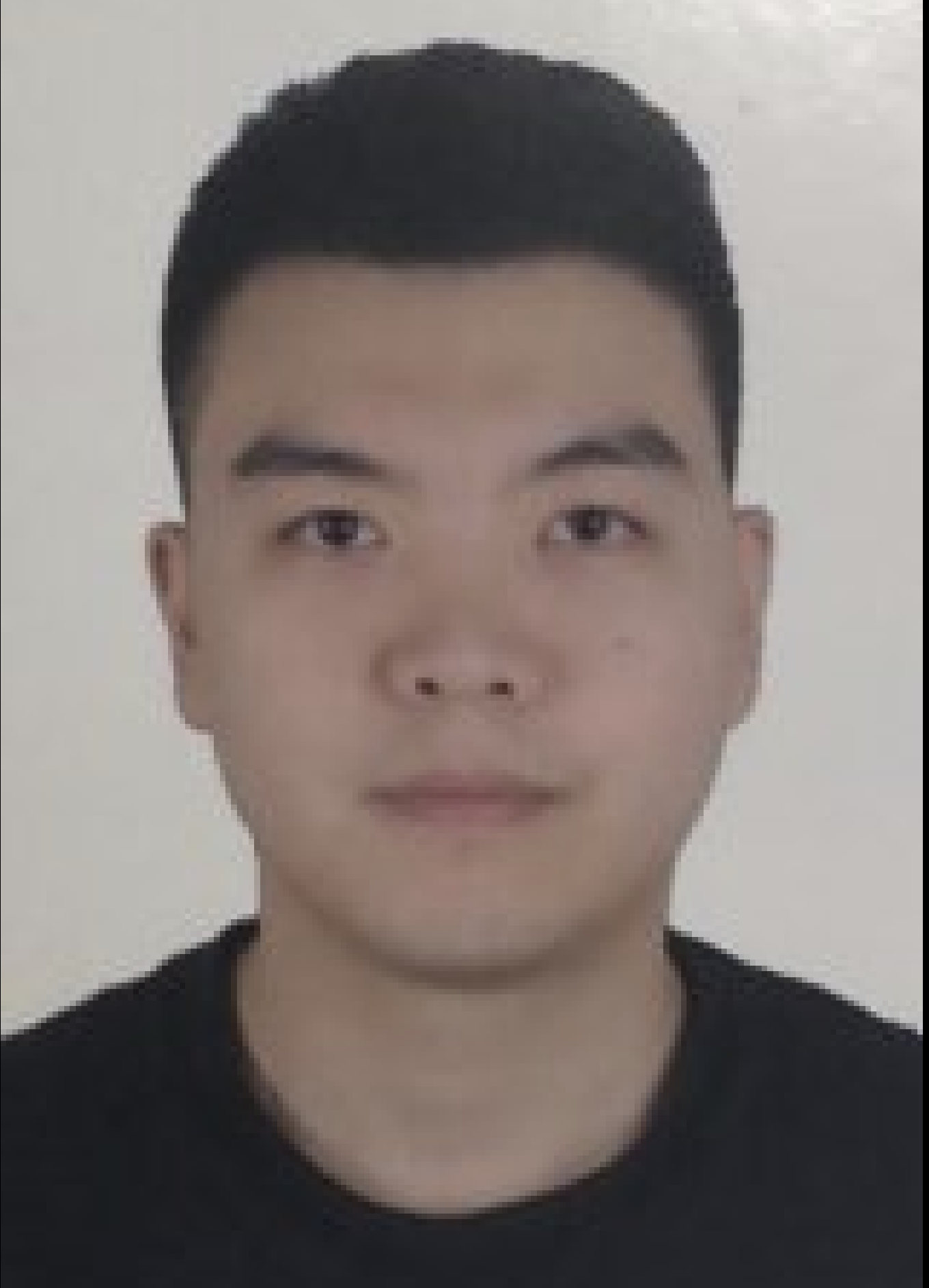}}]{Xiangyu Zhang} (Student Member IEEE) received the B.Sc. degree from the University of Western Australia, Perth, Australia, and the M.Sc. degree from Johns Hopkins University, Baltimore, USA. He is currently pursuing the Ph.D. degree at University of New South Wales, Sydney, Australia, under the supervision of Prof. Julien Epps and Prof. Beena Ahmed. His research interests include speech and language processing, multi-modal learning, foundation models, and digital health.
\end{IEEEbiography}

\begin{IEEEbiography}
[{\includegraphics[width=1in,height=1.25in,clip,keepaspectratio]{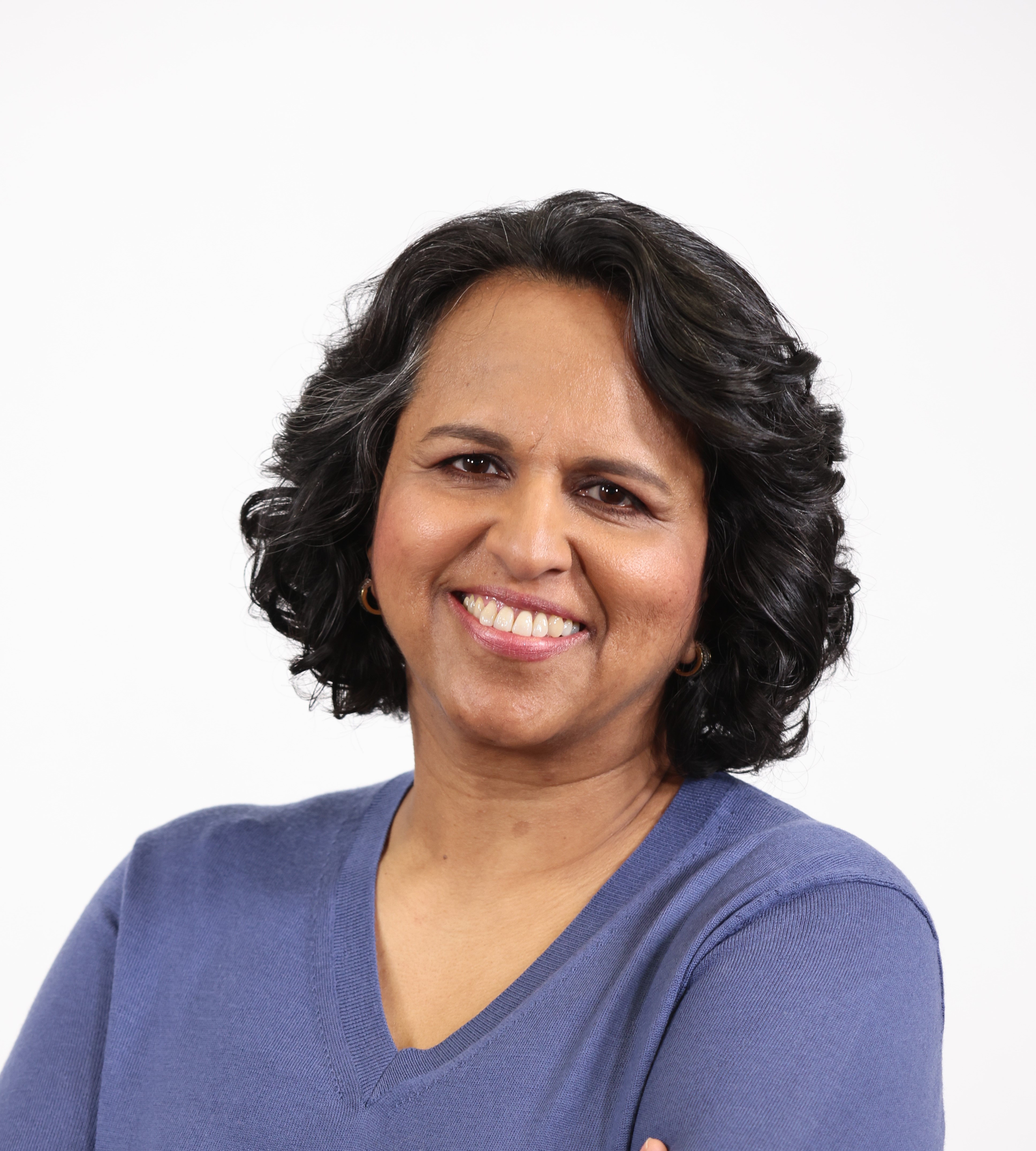}}]{Beena Ahmed} received her Ph.D. degree in Electrical Engineering from the University of New South Wales (UNSW), Sydney, Australia, and B.Sc. (Hons.) degree in Electrical engineering from the University of Engineering and Technology, Lahore, Pakistan, in 2004 and 1993, respectively. She is currently an Associate Professor in Signal Processing at the School of Electrical Engineering and Telecommunications, UNSW, Sydney, Australia. At UNSW, she is the Co-Director of the Signals, Information, and Machine Intelligence Lab and the Technical Lead, Connected Health at the Tyree Foundation Institute of Health Engineering. Prior to that, she was an Assistant Professor in the Department of Electrical and Computer Engineering, Texas A\&M University at Qatar, Doha, Qatar. Her current research interests relate to low resource speech processing areas that includes amongst others disordered speech processing, child speech diarization and recognition, mispronunciation detection and diagnosis and cognitive decline monitoring in speech. Her work has direct practical application in multiple domains such as speech pathology, healthcare and second language learning, and has received multiple awards for translational work. She is a Member of ISCA and general co-chair of Interspeech 2026.
\end{IEEEbiography}

\begin{IEEEbiography}
[{\includegraphics[width=1in,height=1.25in,clip,keepaspectratio]{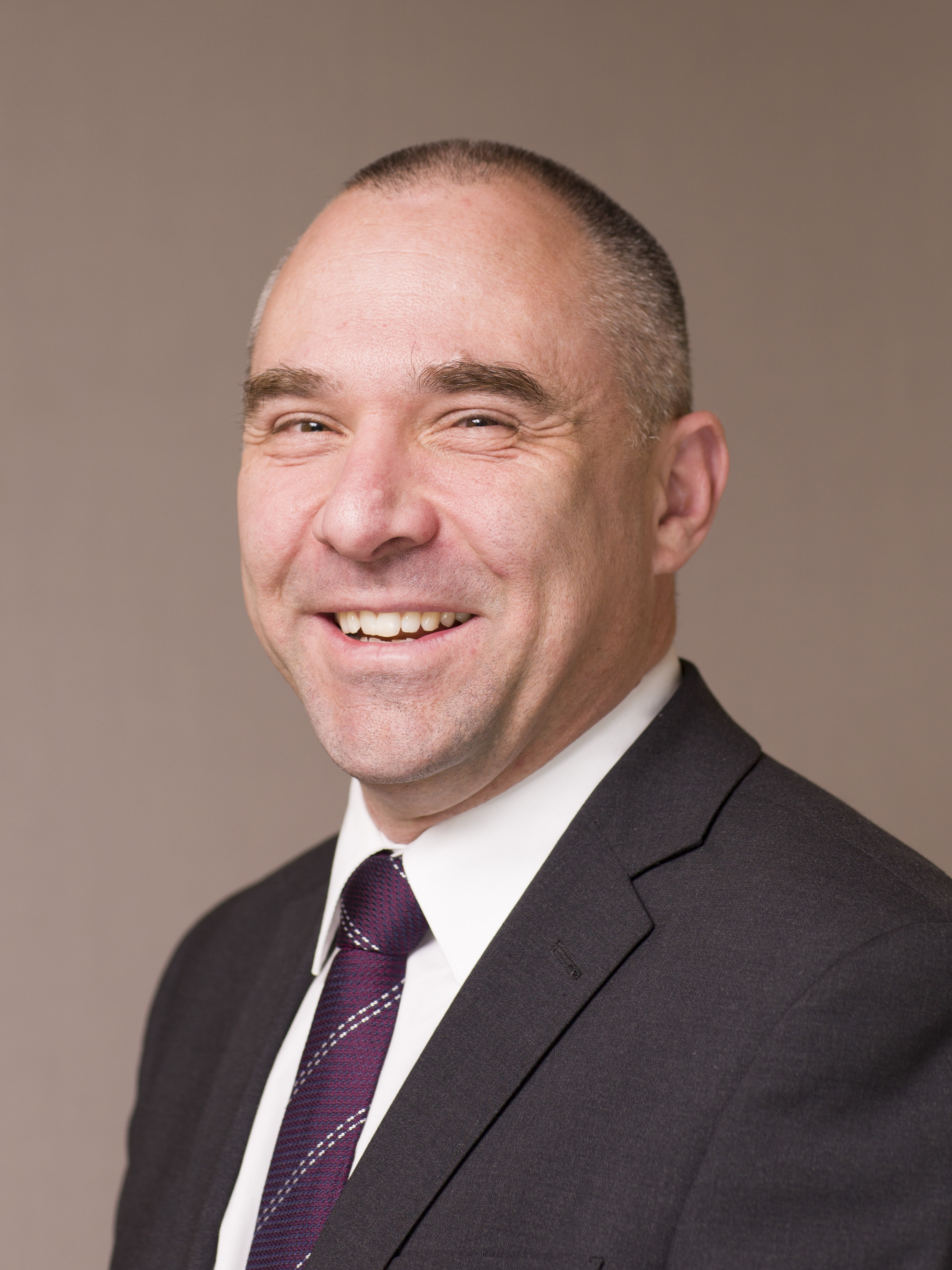}}]
{Julien Epps} (Senior Member, IEEE) received the B.E. and Ph.D. degrees from the University of New South Wales (UNSW), Sydney, Australia, in 1997 and 2001, respectively. From 2002 to 2004, he was a Senior Research Engineer with Motorola Labs, where he was engaged in speech recognition. From 2004 to 2006, he was a Senior Researcher with National ICT Australia, Sydney. He then joined the School of Electrical Engineering and Telecommunications, UNSW, Australia, in 2007 as a Senior Lecturer, and is currently a professor and the Dean of Engineering. He is also a contributed Researcher with Data61, CSIRO, Australia. He has authored or co-authored more than 270 publications and serves as an Associate Editor for \textit{IEEE Transactions on Affective Computing}. His current research interests include characterization, modeling, and classification of mental state from behavioral signals, such as speech, eye activity, and head movement.
\end{IEEEbiography}
\end{document}